\documentclass[pdflatex,twocolumn,sn-nature]{sn-jnl}

\usepackage{geometry}
\usepackage{graphicx}%
\usepackage{svg}
\usepackage{multirow}%
\usepackage{amsmath,amssymb,amsfonts}%
\usepackage{amsthm}%
\usepackage{mathrsfs}%
\usepackage[title]{appendix}%
\usepackage{xcolor}%
\usepackage{textcomp}%
\usepackage{manyfoot}%
\usepackage{subfloat}
\usepackage{booktabs}%
\usepackage{algorithm}%
\usepackage{algorithmicx}%
\usepackage{algpseudocode}%
\usepackage{listings}%
\usepackage{ulem}%
\usepackage{stfloats}%
\usepackage{tabularx}%
\usepackage{array}
\usepackage{float}
\usepackage{tabularx}
\usepackage{makecell}
\usepackage{array} 
\usepackage{svg}
\usepackage{placeins}
\usepackage{wasysym}
\usepackage{caption}

\makeatletter
\renewcommand\abstractname{\large Abstract} 
\renewcommand{\keywordname}{{\normalsize Keywords}}  
\makeatother

\begin{document}



\title[Article Title]{Measuring momentum-resolved dissipation of phonon-polaritons in LiNbO$_3$ with terahertz driving}


\author*[1]{\fnm{Megan F.} \sur{Biggs}}\email{megan.nielson@byu.edu}
\equalcont{These authors contributed equally to this work.}

\author*[2]{\fnm{Rossella} \sur{Acampora}}\email{racampora@phys.ethz.ch}
\equalcont{These authors contributed equally to this work.}

\author*[3]{\fnm{Niccolò} \sur{Sellati}}\email{niccolo.sellati@uniroma1.it}
\equalcont{These authors contributed equally to this work.}

\author[4,5]{\fnm{Mattia} \sur{Udina}}\email{mattia.udina@uniroma1.it}

\author[3]{\fnm{Lara} \sur{Benfatto}}\email{lara.benfatto@roma1.infn.it}

\author[2]{\fnm{Elsa} \sur{Abreu}}\email{elsabreu@pyhs.ethz.ch}

\author[2,6]{\fnm{Steven L.} \sur{Johnson}}\email{johnson@phys.ethz.ch}

\author*[1]{\fnm{Jeremy A.} \sur{Johnson}}\email{jjohnson@chem.byu.edu}

\affil[1]{\orgdiv{Department of Chemistry and Biochemistry}, \orgname{Brigham Young University}, \city{Provo}, \state{UT}, \country{USA}}

\affil[2]{\orgdiv{Institute for Quantum Electronics}, \orgname{ETH Zurich}, \orgaddress{ \city{Zurich}, \postcode{CH-8093}, \country{Switzerland}}}

\affil[3]{\orgdiv{Department of Physics and ISC-CNR}, \orgname{Sapienza University of Rome}, \orgaddress{\street{P.le A. Moro 5}, \city{Rome}, \postcode{00185}, \country{Italy}}}

\affil[4]{\orgdiv{Institut de Physique et Chimie des Matériaux de Strasbourg (UMR 7504)}, \orgname{Université de Strasbourg and CNRS}, \orgaddress{\city{Strasbourg}, \postcode{67200}, \country{France}}}

\affil[5]{\orgdiv{Laboratoire Matériaux et Phénomènes Quantiques}, \orgname{Université Paris Cité, CNRS}, \orgaddress{\city{Paris}, \postcode{75205}, \country{France}}}

\affil[6]{\orgdiv{Center for Photon Science}, \orgname{Paul Scherrer Institut}, \orgaddress{\street{Villigen}, \city{PSI}, \postcode{CH-5232}, \country{Switzerland}}}

%
\abstract{\normalsize Mapping the dispersion of polaritons, hybrid quasiparticles arising from light–matter coupling, can provide key insights into the material dielectric response, coupling strength, and energy transfer pathways with other excitations.
In this work, we present THz pump–Raman probe (TP–RP) as a versatile method for mapping the polariton dispersion in polar non-centrosymmetric materials, demonstrated here for the case of phonon-polaritons in LiNbO$_3$.
By resonantly driving polaritonic modes with a broadband THz pump and probing them with a tunable NIR Raman pulse, TP-RP allows for the extraction of the momentum-dependence of both their frequency and damping rate with high accuracy.
The spectral features observed in the pump--probe signal, including the polaritonic response as well as pulse artifacts, are reproduced within a many-body theoretical approach. Applying the technique to study the \textit{E}(\text{TO}$_1$) phonon of LiNbO$_3$ enables the combined analysis of theory and experiments to uncover a nontrivial frequency dependence of the phonon intrinsic damping rate, revealing possible anharmonic couplings to other modes.}
\keywords{\normalsize Ferroelectrics, Phonon-polaritons, Nonlinear behavior, THz pump–Raman probe, Polaritonic dispersion, Mode coupling, Phonon damping, Propagation effects, Quantum many-body, Phase-matching, LiNbO$_3$}
\maketitle
%
\section*{Introduction}\label{sec1}
Polaritons are hybrid quasiparticles that arise from the coupling between photons and dipolar collective excitations of a material.
Such hybridization gives rise to so-called ``avoided crossings'' in the dispersion at points where the frequencies of the two uncoupled excitations would intersect in momentum space \cite{Low2017,Basov2016,Basov2020,Basov2025}.
Measuring the polariton dispersion in a material gives direct insight into the dynamics of the modes and the magnitude of light–matter coupling \cite{Kuzmenko2005,Basov2025} and provides indirect information on the interactions with other excitations through the damping rate of the polaritonic modes \cite{Caldwell2015,Kansanen2019}.

In this paper, we focus on phonon-polaritons.
These light-matter modes form through the hybridization between photons and transverse-optical (TO) IR-active phonons in polar dielectrics, and they have been extensively studied since their discovery in the mid-20th century \cite{feurer_review07}. 
More recently, several studies have demonstrated how the coupling between light and IR-active phonons plays a key role in the ultrafast control of materials. Large mode displacements driven by THz fields are thought to enable processes such as polarization switching in ferroelectric states \cite{Kampfrath2013} and induced chiral states in antiferrochiral materials \cite{Zeng2025}, highlighting the growing interest in hybrid light–matter excitations.

Inelastic neutron and X-ray scattering are the standard techniques for measuring phonon dispersion relations. These methods provide access to the entire Brillouin zone with momentum resolution of $10^{-2}$ to $10^{-3}$ \r{A}$^{-1}$ and energy resolution ranging from the $\text{meV}$ to the $\mu\text{eV}$ scale \cite{Baron2020,Shirane2002,Ollivier2004,Lory2017}.
Optical probes complement these methods by resolving features near the $\Gamma$ point with a momentum resolution of $10^{-6}$ \r{A}$^{-1}$ and sub$-\text{meV}$ energy resolution \cite{Knighton2018,Luo2024}, making them particularly well suited for investigating the photon--phonon crossing.
Common optical probes for mapping phonon-polaritons that are simultaneously IR- and Raman-active typically rely on conventional equilibrium spectroscopies such as spontaneous Raman scattering \cite{Henry1965}. These methods, however, suffer from weak signal intensity and limited momentum coverage.
Alternative nonlinear techniques such as Impulsive Stimulated Raman Scattering (ISRS) \cite{Planken1992,Crimmins2002,Merlin2003,Knighton2018} overcome these limitations by providing stronger signals and access to a broader range of momenta of the polariton branches, through a pump--probe interaction scheme that relies on optical pulses  \cite{Knighton2018}. 
Like spontaneous Raman scattering, ISRS uses high-frequency light to probe the dispersion. This restricts the technique to materials with a bandgap larger than the pump photon energy, and also limits the ability to enhance signal levels by increasing the pulse intensity since one should avoid the regime where multiphoton absorption cross-sections become non-negligible. 
Moreover, collinear ISRS probes forward- and backward-propagating polaritons simultaneously, leading to overlapping spectral peaks and thereby increasing the uncertainty in the extracted frequencies and damping rates \cite{Knighton2018}.

Advances in the generation of intense and phase-stable terahertz (THz) pulses allow new opportunities for the excitation and detection of low-energy collective modes in polar materials \cite{THzReview2023,THzReview2025}. 
Broadband THz pulses can span a wide frequency range \cite{THzReview2023}, 
and can simultaneously excite multiple vibrational polaritonic modes within the pump bandwidth, as demonstrated in lithium niobate (LiNbO$_3$) \cite{Dastrup2017,Knighton2019,Blake2022}. 
Recently, Luo et al.\ also demonstrated how THz-field-induced second-harmonic generation can be used to map the phonon-polariton dispersion in van der Waals materials \cite{Luo2024}.
These works highlight the potential of THz-driven approaches for probing polaritonic properties, opening the door to a largely unexplored landscape of THz-based techniques.
In this work, we introduce a new method to map the dispersion in non-centrosymmetric samples using a THz pump-Raman Probe (TP-RP) transmission configuration, and we demonstrate it experimentally on ferroelectric LiNbO$_3$, where an anomalous behavior of the polariton damping rate has been previously reported \cite{Schwarz1996,Crimmins2002,Knighton2018}.

LiNbO$_3$ is a negative uniaxial wide-bandgap insulator \cite{Weis1985}
with a large second-order susceptibility \cite{Dhar1990}, widely used to generate high-field THz pulses \cite{Hirori2011,Zhang2012,Zhang2021}.
Its primitive unit cell has a rhombohedral structure with $R3c$ symmetry and contains 10 atoms, resulting in 27 optical phonon branches.
Our TP-RP approach makes use of a linearly polarized broadband THz pump to directly excite polaritons derived from $E$-symmetry TO phonons of LiNbO$_3$. In particular, the dispersion of the lowest-energy $E(\text{TO}_1)$ mode \cite{Knighton2018} falls entirely within the bandwidth of our THz pulse, providing an ideal testbed for the technique.
The dynamics of the driven phonon-polariton are probed by measuring the changes in the polarization of a femtosecond near-infrared (NIR) pulse with a tunable central frequency. Phase-matching of the variable-wavelength NIR probe to the hybrid mode provides the momentum selectivity required to map out the momentum-dependent frequency and damping rate.

For a theoretical analysis of the experimental findings, we build on the results of Ref.\ \cite{Sellati2025} and develop a many-body framework to describe the second-order TP-RP process in uniaxial crystals, and we 
provide a systematic treatment of propagation effects to connect the internal material response to the measured pump--probe signal.
The combined experimental and theoretical results enable a straightforward reconstruction of the dispersion of the $E(\text{TO}_1)$ phonon-polariton with reduced uncertainty compared to collinear ISRS \cite{Knighton2018}. Moreover, we are able to extract the frequency dependence of the intrinsic phonon damping rate in a region not accessible by linear spectroscopy, and reveal anharmonic couplings responsible for the anomalous behavior of the polariton damping rate \cite{Wiederrecht1995,Schwarz1996,Crimmins2002,Knighton2018}.
Our approach is not restricted to phonon-polaritons in LiNbO$_3$, but can be extended to other non-centrosymmetric polaritonic systems, provided that phase-matching conditions can be realized.
%
%
\section*{Experiment}\label{sec2}
\subsection*{Results}
\begin{figure*}[t]
    \centering
    \includegraphics[width=1.0\linewidth]{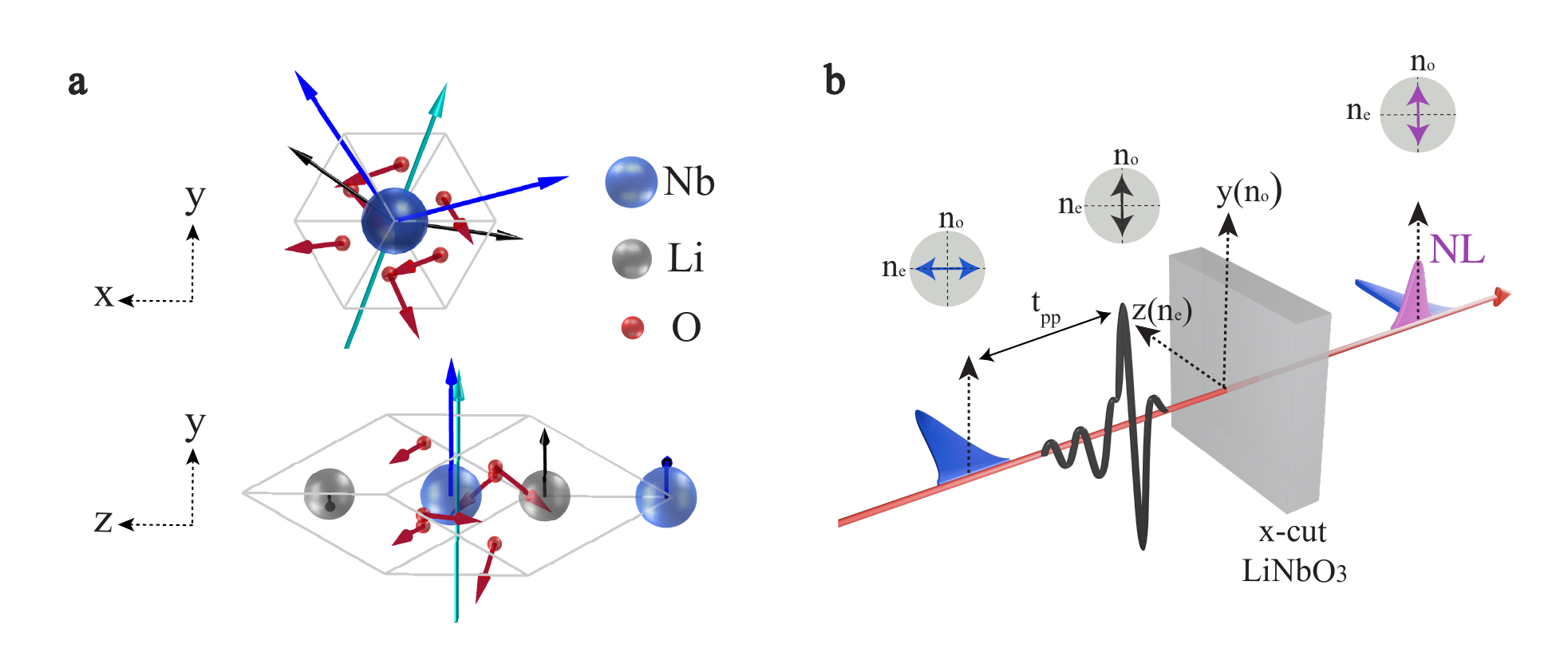}
    \caption{\textbf{LiNbO$_3$ unit cell and pump--probe scheme.} \textbf{a} Rhombohedral unit cell of LiNbO$_3$ with the DFT-calculated eigenvector of the THz-driven E(TO$_1$) phonon-polariton mode shown as arrows indicating the displacement directions of the Nb (blue), Li (black) and O (red) atoms. 
    The dipole moment associated with this mode is represented with cyan arrows and lies in the xy plane.
    \textbf{b} Schematic of the THz pump--Raman probe experiment on a 500 $\mu\text m$-thick x-cut LiNbO$_3$ crystal, showing the pulse sequence, incidence angle on the sample, and polarization configuration. Gray circles indicate the polarization of the incident probe (blue) along the extraordinary ($e$) axis, corresponding to the crystallographic z, and of the pump (gray) along the ordinary ($o$) axis, corresponding to y. The NIR probe field propagates through the crystal and generates the nonlinear (NL) signal (purple), polarized along the ordinary y axis.}
    \label{}
\end{figure*}
\begin{figure*}[t]
    \centering
    \includegraphics[width=1\linewidth]{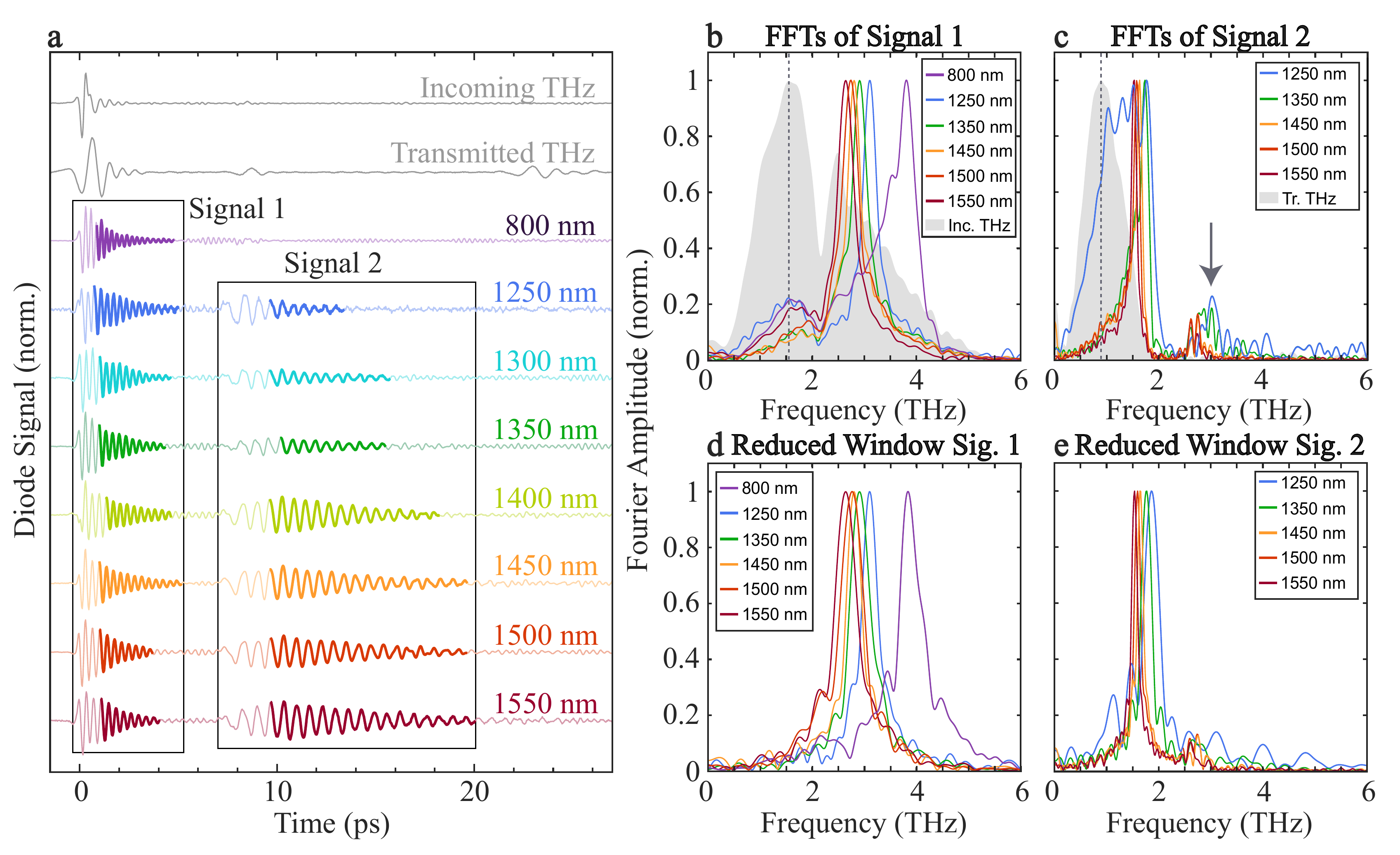}
    \caption{\textbf{Experimental spectra with varying probe wavelength.} \textbf{a} Time traces of the THz pump field generated with a NIR pump at 1450 nm, THz field transmitted through the LiNbO$_3$ sample, and nonlinear signals corresponding to different probe wavelengths. Time-separated contributions Signal 1 and Signal 2 are indicated by black boxes. \textbf{b,c} Full time-domain FFT of Signal 1 (b) and Signal 2 (c) for selected probe wavelengths. Vertical dashed gray lines highlight the main-peak frequency of the spectrum of the incident (b) or transmitted (c) THz field (gray area). The gray arrow indicates satellite features in Signal 2 associated with residual oscillations of Signal 1. \textbf{d,e} FFT of Signal 1 (d) and Signal 2 (e) for selected probe wavelengths, performed over the reduced time-windows indicated by the darker lines in panel a, which exclude the temporal overlap with the THz pulse.}
    \label{expResults}
\end{figure*}
The $E(\text{TO}_1)$ phonon-polariton mode of LiNbO$_3$ is associated with a dipole moment lying in the crystallographic xy plane \cite{Parlinski2000}, as shown in Fig.\ \ref{}a.
To probe the momentum-resolved dynamics of the phonon-polariton associated with this mode, we apply the TP-RP technique to a 500 $\mu\text m$-thick x-cut crystal, with both pump and probe fields at normal incidence on the sample. 
The pulse sequence and polarizations are illustrated in Fig.\ \ref{}b.
We generate an intense THz field through optical rectification of NIR pulses in a PNPA crystal \cite{Rader2022} (Appendix \ref{field}) and polarize the THz electric field along the crystallographic y axis of LiNbO$_3$, so that it can resonantly drive the $E$ mode (Appendix \ref{symmetry}). 
The Raman detection is performed using a collinear optical probe (see Methods) polarized along the z axis, with wavelength tuned between 800 nm and 1550 nm. In the following we refer to the propagation direction of the incident probe as the forward direction.
Because the Raman tensor associated with $E$-modes in LiNbO$_3$ has nonvanishing off-diagonal components \cite{Sanna2015}, the probe scattered by the nonlinear process has a polarization that is perpendicular to the incident probe polarization.
The phase-delay between the unscattered light and the scattered probe leads to an overall elliptical probe polarization.
We then track the pump--probe dynamics by measuring the probe ellipticity changes through balanced detection as a function of the time delay between the two pulses (see Methods), and we measure the sample response over a $30 \text{ ps}$ time window.

The time evolution of the measured pump–probe signal for all investigated probe wavelengths is shown in Fig. \ref{expResults}a, 
together with the incident and transmitted THz waveforms (gray traces in Fig.\ \ref{expResults}a) measured with electro-optic sampling (Appendix \ref{field}).  
With the exception of the data collected using a probe wavelength of 800 nm, the response consists of two oscillating contributions that are separated in time, denoted Signal 1 and Signal 2. The time traces are normalized to the maxima of Signal 1. Both signals oscillate around zero. 
Signal 1 is characterized by higher-frequency oscillations, starting at zero pump--probe delay and persisting up to $5 \text{ ps}$. This signal is generated by the incident pump field, traveling forward in the sample.
Signal 2 consists of lower-frequency oscillations emerging after approximately
$7\text{ ps}$, consistent with the delay expected from THz propagation through the 500 $\mu\text{m}$ sample (Appendix \ref{timesep}), and lasting approximately 10 ps. This signal is generated by the pump field back-reflected at the rear interface, traveling backwards in the material.
For both signals, the oscillation frequencies and decay times display a clear dependence on the probe wavelength.

We compute the fast Fourier transform (FFT) of the pump--probe signal in time windows from approximately $-1$ ps to $5$ ps to isolate Signal 1, and from approximately $7$ ps and $20$ ps to isolate Signal 2. We plot their normalized spectral amplitudes for a representative subset of probe wavelengths in Fig. \ref{expResults}b and Fig. \ref{expResults}c, respectively.
The complete datasets and the precise FFT time windows for every measured signal are reported in Appendix \ref{ffts}.
The spectra of the incident and transmitted THz fields generated with a NIR pump at 1450 nm are also reported as gray shaded areas in Figs. \ref{expResults}b and \ref{expResults}c. As discussed below, the nonlinear signal falls entirely within the pump spectrum due to the direct THz excitation of the phonon mode. The transmitted field lacks components above 2 THz due to strong absorption within LiNbO$_3$. The backscattered pump pulse at the rear interface of the sample, involved in the generation of Signal 2, is expected to display the same spectral content of the transmitted THz field.

Both signals exhibit a clear peak that shifts to higher frequency with decreasing wavelength: in Signal 1, the peak frequency shifts from 2.64 THz (1550 nm) to 3.82 THz (800 nm); in Signal 2 it shifts from 1.53 THz (1550 nm) to 1.77 THz (1250 nm) (Appendix \ref{ffts}).
Additionally, broad spectral shoulders appear at frequencies corresponding to the main peak of the THz field, marked by gray dashed lines in Figs. \ref{expResults}b and \ref{expResults}c. These features originate from the driven response of the IR-active mode during the temporal overlap of the pump and probe pulses.
Signal 2 exhibits additional high-frequency peaks, indicated by the gray arrow. These features are observed at the same frequencies as the main peaks of Signal 1 and display a similar dependence on the probe wavelength, indicating that Signal 1 has not fully decayed in the Signal 2 temporal window. 

To suppress features arising from the THz driving field, and isolate the polaritonic contribution, we also perform the FFT over the reduced time-window indicated by the thicker lines in the time traces in Fig.\ \ref{expResults}a, excluding the portions of Signal 1 and Signal 2 that overlap with the THz pulse in time (2 ps).
The resulting normalized spectra are shown in Figs.\ \ref{expResults}d and \ref{expResults}e.
Removing the pump-driven region of the signal allows for a more accurate extraction of the peak frequencies, as well as the inverse-lifetime of the phonon-polariton, determined with the peak full-width at half maximum (FWHM).

\subsection*{Discussion}
\begin{figure*}[t]
    \centering
    \includegraphics[width=0.9\linewidth]{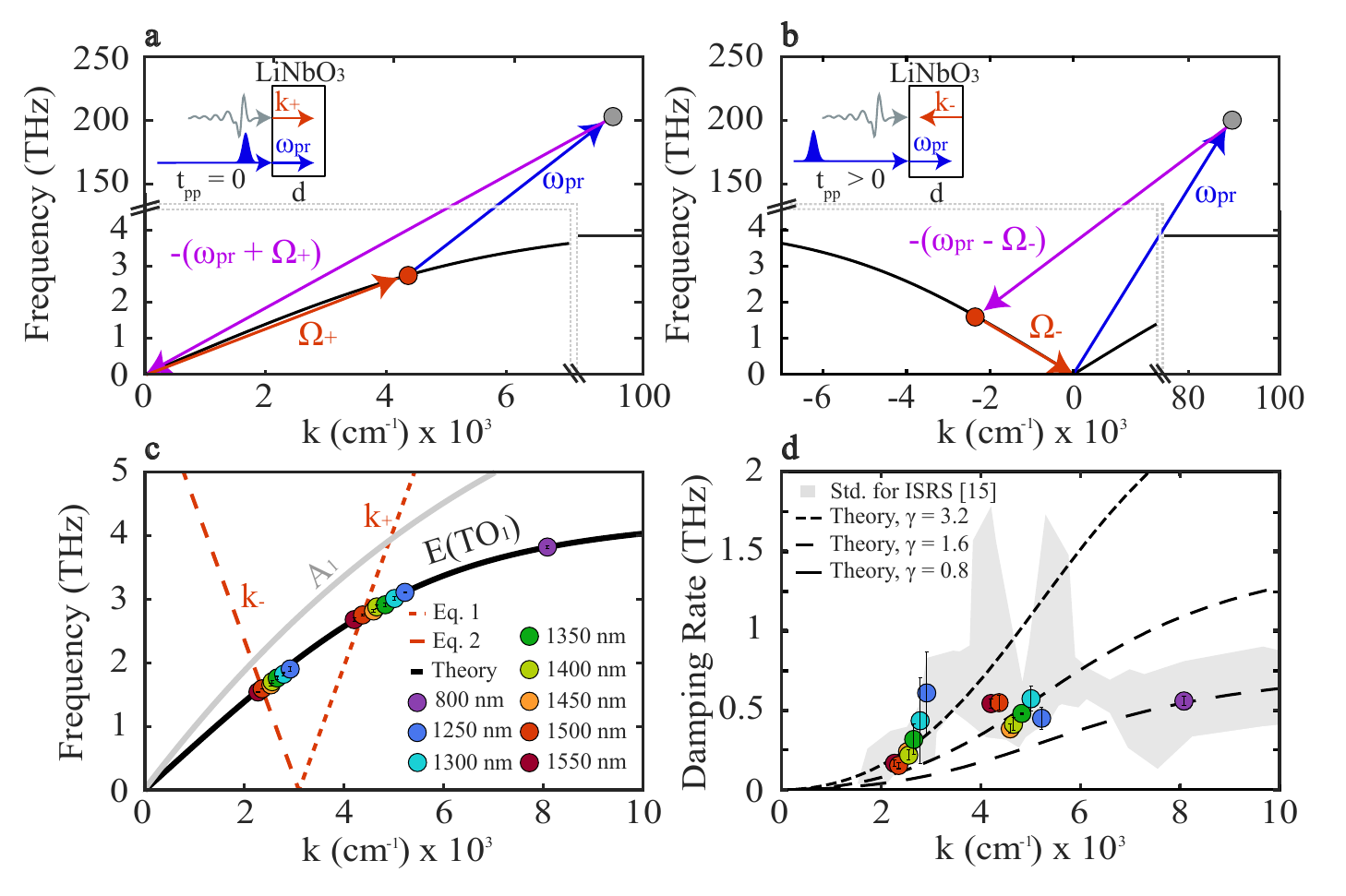}
    \caption{\textbf{Reconstruction of the polaritonic dispersion.} \textbf{a,b} Phase-matching diagrams for forward (a) and backward (b) propagating THz pump, for a 1500 nm monochromatic probe at frequency $\omega_\text{pr}$. The pump interaction is a one-photon direct-excitation process, where forward-propagating polaritons are launched by the incident THz field (inset, a), and backward-propagating polaritons are launched by the THz pulse reflected at the rear interface of the sample (inset, b).  
    The three-wave-mixing scattering can be seen as an SFG or DFG process, respectively, where a probe photon (blue vector) at $\omega_{\mathrm{pr}}$ and a polariton (red vector) produce scattered photons (purple vector) at frequency $\omega_{\mathrm{pr}}+\Omega_{+}$ and $\omega_{\mathrm{pr}}-\Omega_{-}$. Phase-matching allows for the extraction of two points (red dots) along the dispersion (black curve) for each measurement. The vectors are shown as continuous lines for visual clarity, even though they would appear discontinuous due to the axis breaks.
    \textbf{c,d} Reconstructed real (c) and imaginary (d) part of the $E(\text{TO}_1)$ phonon-polariton dispersion. Experimental points are compared with the dispersion found from Eq.\ \eqref{wave_eq} (black curves), with $\varepsilon_\infty=22.47$, $\omega_\text{TO}/2\pi=4.44$ THz, $\omega_\text{LO}/2\pi=5.94$ THz, and three different constant values for $\gamma$ (3.2 THz, 1.6 THz, 0.8 THz). The experimental error bars in panels c and d are calculated as the standard deviation of independent datasets (Appendix \ref{ffts}). Red dashed lines in panel c represent Eqs.\ \eqref{phase_matching1} and \eqref{phase_matching2} for the 1500 nm probe, whose intersections with the dispersion identify the phase-matched points.
    For reference, the expected dispersion curve of the $A_1$ phonon-polariton, as reported in Ref.\ \cite{Blake2022}, is also shown as a gray curve in panel c. The gray shaded area in panel d represents the phonon-polariton damping rate, including experimental uncertainty, measured with ISRS in Ref.\ \cite{Knighton2018}.}
    \label{expDiscussion}
\end{figure*}
To understand the phonon-polariton pump--probe mechanism, it is useful to treat the pumping and probing processes separately. In the following, quantities denoted by $\omega$ or $\Omega$ refer to angular frequencies. For consistency with the measurements, the plots report the corresponding cycle frequencies.

TP-RP relies on a one-photon direct driving mechanism, making the pumping process intrinsically directional: the THz pump launches polaritons propagating exclusively along its propagation direction.
Specifically, the forward-going THz field drives forward-propagating phonon-polaritons, which are characterized by a positive wavevector $k_+$,
as sketched in the inset of Fig.\ \ref{expDiscussion}a. The attenuated THz pulse reflected at the rear interface of the sample launches backward-propagating polaritons, characterized by negative $k_-$, as illustrated in the inset of Fig.\ \ref{expDiscussion}b. 
The finite propagation time of the THz light through the thick sample causes the clear temporal separation between Signal 1 and Signal 2 in the measurements.

In the TP-RP probing process, detection relies on Raman scattering from the driven mode via three-wave-mixing. Specifically,
coherent phonon-polaritons at frequency $\Omega_\pm$ travel inside the material and mix with probe photons at frequency $\omega_\text{pr}$.
The nonlinear scattering can occur via sum-frequency generation (SFG) for forward-propagating polaritons, corresponding to an anti-Stokes process, or via difference-frequency generation (DFG) for backward-propagating modes, corresponding to a Stokes process \cite{Ikegaya2015}.
NIR photons are emitted during the scattering event, with frequency $\omega_\text{pr} + \Omega_{+}$ for SFG and $\omega_\text{pr} - \Omega_{-}$ for DFG. 
In a time-resolved pump–probe experiment, the measured signal is the phonon-induced relative change in the probe pulse intensity, which encodes the interference between optical photons at the probe frequency $\omega_\text{pr}$ and scattered photons shifted by $\pm\Omega_\pm$ \cite{Udina2019}. As a result, although the emitted photons have optical frequencies, the detected intensity exhibits oscillations at
$\Omega_\pm$, arising from the beating between $\omega_\text{pr}$ and $\omega_\text{pr} \pm \Omega_{\pm}$.

In principle, the broadband THz field drives any frequency within its bandwidth, meaning that the phonon-polariton involved in the three-photon scattering can have any frequency within the pump spectrum. In contrast, the probe interaction is most efficient for polariton frequencies $\Omega_\pm$ that satisfy the phase-matching condition $ k_\pm+k_\text{in}-k_\text{out}=0$, corresponding to momentum conservation in the three-wave-mixing process.
Here, $k_\pm$ represent the wavevectors of the forward ($+$) or backward ($-$) propagating THz polariton, and $k_\text{in}$ and $k_\text{out}$ are the momenta of incident and scattered optical light, respectively polarized along the extraordinary ($e$) axis, corresponding to the crystallographic z, and the ordinary ($o$) axis, corresponding to y. We schematically illustrate the SFG and DFG processes for a 1500 nm probe ($\omega_\text{pr}/2\pi=200\text{ THz}$) in Figs.\ \ref{expDiscussion}a and \ref{expDiscussion}b, respectively. In both panels, the blue vector represents the incoming probe photon, the purple vector is the emitted photon, and the red vector indicates the phase-matched point $(k_\pm,\Omega_\pm)$ along the polariton dispersion curve, shown in black.

The phase-matching condition can be separated into two equations for forward- and backward-propagating modes,
\begin{align}
    k_+&=-\frac{n_e}{c}\omega_\text{pr}+\frac{n_o}{c}(\omega_\text{pr}+\Omega_+),\label{phase_matching1}\\
    k_-&=\frac{n_e}{c}\omega_\text{pr}-\frac{n_o}{c}\big(\omega_\text{pr}-\Omega_-\big)\label{phase_matching2},
\end{align}
when a positive probe frequency $\omega_\text{pr}$ is used. We note that, for a positive $\omega_\text{pr}$, the red arrow in Fig.\ \ref{expDiscussion}b would point up. In Eqs. \eqref{phase_matching1} and \eqref{phase_matching2},
$n_o$ and $n_e$ are the ordinary and extraordinary optical refractive indices at a given probe wavelength, and $c$ is the speed of light in vacuum.
As a consequence of phase-matching imposed by propagation in the medium, the detection process selectively enhances specific combinations of frequencies and momenta that satisfy Eqs. \eqref{phase_matching1} and \eqref{phase_matching2} for forward- or backward-propagating polaritons, respectively, while suppressing scattering pathways with large wavevector mismatch.
From the FFT of the experimental signal obtained using the reduced time windows, Figs.\ \ref{expResults}d and \ref{expResults}e, we can directly extract the phase-matched frequencies
$\Omega_+$ and $\Omega_-$ for each probe wavelength as the ones corresponding to the main peaks of Signal 1 and Signal 2, respectively. Then Eqs.\ \eqref{phase_matching1} and \eqref{phase_matching2} are used to determine the momenta $k_+$ and $k_-$ of the scattered polaritons.

By varying the probe wavelength, we obtain 15 phase-matched points, two for each wavelength except at 800 nm. In this case, Signal 2 is not observed because its expected phase-matched frequency, near 3 THz, is strongly absorbed by the sample and thus missing in the internally-reflected THz pulse (see Transmitted THz spectrum in Fig.\ \ref{expResults}c).
The experimental phase-matched points can be used to reconstruct the frequency-momentum dispersion relation of the $E$(TO$_1$) phonon-polariton in LiNbO$_3$, as shown in Fig.\ \ref{expDiscussion}c.
Analogously, we can extract the imaginary part of the dispersion (Fig.\ \ref{expDiscussion}d), that represents the momentum-dependence of the phonon-polariton damping rate, from the FWHM of the experimental peaks.

We compare the extracted points with the real and imaginary parts of the low-energy branch of the polariton dispersion relation \cite{Dougherty1992}, found as the lower-frequency solution of the wave-equation 
\begin{align}\label{wave_eq}
n_\text{THz}^2(\Omega)\,\Omega^2-c^2k^2=0.
\end{align} 
Here,
\begin{align}\label{nTHz}
    n_\text{THz}(\Omega)=\sqrt{\varepsilon_\infty\frac{\Omega^2+2i\gamma\Omega-\omega_\text{LO}^2}{\Omega^2+2i\gamma\Omega-\omega_\text{TO}^2}}
\end{align}
is the refractive index at THz frequencies obtained from a single-Lorentz oscillator model, that accounts for the coupling of THz light to the TO phonon. The quantities $\omega_\text{TO}$ and $\omega_\text{LO}$ are the transverse-optical and longitudinal-optical phonon frequencies, respectively, $\varepsilon_\infty$ is the high-frequency electronic dielectric constant, and $\gamma$ is the intrinsic bare-phonon damping rate. 
Eq.\ \eqref{wave_eq} is solved numerically for the complex $\Omega$ as a function of $k$, and we only consider the lower-frequency solution. Its real and imaginary parts are shown as black curves in Figs.\ \ref{expDiscussion}c and \ref{expDiscussion}d. The real part is divided by $2\pi$ to obtain the cycle frequency, whereas the imaginary part is left unchanged to represent the true inverse polariton lifetime.
Values for $\omega_\text{TO}$, $\omega_\text{LO}$ and $\varepsilon_\infty$ are taken from Ref.\ \cite{Knighton2018}, while three different constant values of $\gamma$ are considered, ranging from low to high damping rates.
Red dashed lines, corresponding to Eqs.\ \eqref{phase_matching1} and \eqref{phase_matching2} for a 1500 nm probe, are shown in Fig.\ \ref{expDiscussion}c, and their intersections with the black curve indicate the phase-matched points along the dispersion.

The real part of the dispersion (Fig.\ \ref{expDiscussion}c) is in excellent agreement with the experimental data,
confirming the consistency of TP-RP measurements with the expected polariton dispersion in uniaxial LiNbO$_3$. In contrast, the imaginary part (Fig.\ \ref{expDiscussion}d) cannot be reproduced by the theoretical dispersion for any constant value of $\gamma$, consistent with previous observations of an anomalous behavior \cite{Wiederrecht1995,Schwarz1996,Crimmins2002,Knighton2018}. In particular, the experimental values of the phonon-polariton damping rate obtained by TP-RP agree with those measured by ISRS in Ref.\ \cite{Knighton2018}, for which the standard deviation is shown as a gray shaded area in Fig.\ \ref{expDiscussion}d. 

Our TP-RP method enables the measurement of the real and imaginary parts of the dispersion with improved accuracy and smaller error bars compared to all-optical setups, as evidenced by the comparison with ISRS in Fig.\ \ref{expDiscussion}d.
This stems from the temporal separation between Signal 1 and Signal 2. 
Such an improvement is essential when subtle variations in the damping rate need to be resolved with high accuracy, as is expected in more complex systems where additional features may emerge from coupling to other modes \cite{Dougherty1992}, field-induced band splitting \cite{Yaniv2025}, or pulse propagation away from a principal axis. The signal separation opens promising possibilities for extending the TP-RP technique to more advanced spectroscopies, such as 2D-THz spectroscopy \cite{Knighton2019,Blake2022,Johnson2019}, where it could help disentangle different mode contributions.
Another advantage of this experimental technique is that, by tailoring the pump spectrum, higher-frequency branches such as the $E$(TO$_2$) and $E$(TO$_3$) modes can also be probed \cite{Blake2022}.
More generally, access to different dispersion branches relies on the possibility of achieving phase-matching.
For example, the lowest $A_1$-symmetry phonon-polariton branch, shown in gray in Fig. \ref{expDiscussion}c, is not accessible in our experiment despite lying within the THz pump bandwidth, because phase-matching cannot be realized, as discussed in Appendix \ref{phasematching}.
Other $A_1$ modes could instead be accessed by employing a pump with sufficiently high frequency, since the shape of their branches allows phase-matching \cite{Kojima2018B}. 

We note that Eq.\ \eqref{nTHz} is valid only for a single Lorentz oscillator, and, in the case of LiNbO$_3$, it can be used to describe only the lowest phonon-polariton branch. 
We also note that Eqs.\ \eqref{phase_matching1} and \eqref{phase_matching2} are reliable when a narrowband probe field is used, such as the one employed in our experiment ($\pm 10$ nm), for which the resulting spread in phase-matched frequencies remains small (Appendix \ref{ffts}).
To extend our technique to more general experimental conditions, including large probe bandwidths, we introduce a comprehensive theoretical framework that is able to disentangle the various broadening effects such as the finite distribution of phase-matched frequencies, the THz pump bandwidth, and the phonon damping rate $\gamma$.
%
%
\section*{Theory}
\subsection*{Model}
We model the pump–probe interaction as a three-wave-mixing scattering process, mediated by the bare $E(\text{TO}_1)$ phonon, which is simultaneously IR- and Raman-active. Details of the formalism, here generalized to uniaxial crystals from Ref.\ \cite{Sellati2025}, are given in Appendix \ref{theory}.
In a balanced-detection scheme, the measured differential signal is proportional to the nonlinear signal $\text{E}_\text{NL}$ \cite{Basini2024,Sellati2026}, which scales linearly in the pump and probe fields. 
In our theoretical approach, we describe the internal response of the system through a frequency-dependent second-order nonlinear current $\text{J}^{(2)}$ generated inside the material. Following standard practice in electronic systems, we express the current in response to an applied gauge potential $\text{A}$, and compute it using a many-body diagrammatic framework. 
The current acts as a source for the generated signal $\text{E}_\text{NL}$, which travels through the sample while being shaped by the linear optical response of the medium, multiple internal reflections, and finally transmission at the rear interface. These propagation effects are taken into account with a generalized Maxwell-Fresnel approach for collinear pulses \cite{Huber2021,Sellati2025,Fiore2024,Fiore2026}. In the following, all spatial indices of the electromagnetic and phonon fields are suppressed for simplicity.

The transmitted nonlinear field contains all possible combinations of forward- ($t$) and backward- ($r$) propagating pump ($\text A_\text{p}$) and probe ($\text A_\text{pr}$) fields, where the forward direction is defined by the propagation direction of the incident experimental pulses.
To individually analyze Signal 1 and Signal 2, we combine the forward-propagating probe with forward- or backward-propagating pump fields, respectively, and we decompose the nonlinear signal $\text{E}_\text{NL}$ into two contributions, $\text E_\text{NL}^{(1)}$ and $\text E_\text{NL}^{(2)}$.
The first term originates from a forward-propagating pump field $\text A_\text{p}^t$, and it is given by
\begin{align}\label{for_prop}
    \text E_\text{NL}^{(1)}(\Omega)\!\propto\!\int\!d\omega^\prime t_o(\omega^\prime )\frac{\text A^t_{\text p}(\Omega)\text K^{(2)}(\omega^\prime,\Omega)\text A^t_{\text{pr}}(\omega^\prime\!-\!\Omega)}{\frac{n_\text{THz}(\Omega)}{c}\Omega+\frac{n_e}{c}(\omega^\prime\!-\!\Omega)-\frac{n_o}{c}\omega^\prime }.
\end{align}
The second term is associated with a backward-propagating pump field $\text A_\text{p}^r$, and it is given by
\begin{align}\label{back_prop}
    \text E_\text{NL}^{(2)}(\Omega)\propto\!\int\!d\omega^\prime  t_o(\omega^\prime )\frac{\text A^r_{\text p}(\Omega)\text K^{(2)}(\omega^\prime,\Omega)\text A^t_{\text{pr}}(\omega^\prime\!-\!\Omega)}{-\frac{n_\text{THz}(\Omega)}{c}\Omega\!+\!\frac{n_e}{c}(\omega^\prime\!-\!\Omega)\!-\!\frac{n_o}{c}\omega^\prime }.
\end{align}
Here, $\omega^\prime$ denotes the frequency of the scattered optical signal modulated by the ordinary-axis transmission coefficient $t_o(\omega^\prime)$.
In both equations, the numerators describe the microscopic light–matter interaction, where the internal second-order response is captured by the interaction kernel,
\begin{align}\label{kernel}
\text K^{(2)}(\omega^\prime,\Omega)\propto\frac{Z^*R}{\Omega^2+2i\gamma\Omega-\omega_\text{TO}^2}.
\end{align}
The kernel is proportional to the second-order susceptibility and contains the bare-phonon propagator, with an overall strength set by the mode effective charge $Z^*$ and the phonon Raman tensor $R$.
The denominators of Eqs.\ \eqref{for_prop} and \eqref{back_prop} express the phase-matching equations, and determine the conditions under which the nonlinear interaction contributes efficiently to the observed signal. Here, the wavevectors of the propagating THz pulses are expressed as $k_\pm=\pm n_\text{THz}(\Omega)\Omega/c$. The phase-matching conditions Eqs. \eqref{phase_matching1} and \eqref{phase_matching2} are recovered for a monochromatic probe field $\text{A}_\text{pr}(\omega)\propto \big(\delta(\omega-\omega_\text{pr})+\delta(\omega+\omega_\text{pr})\big)$.

\begin{figure*}[t]
    \centering
    \includegraphics[width=0.9\textwidth]{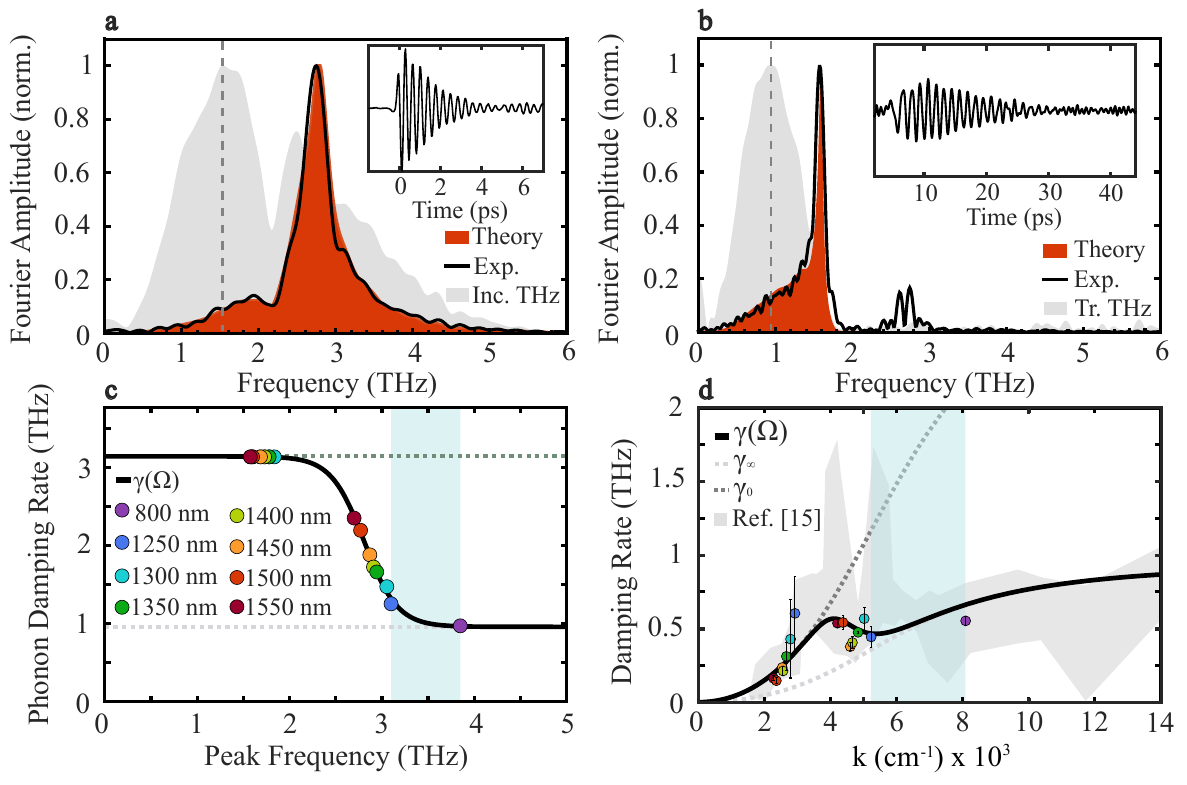}
    \caption{\textbf{Theoretical pump--probe response and damping rate.} \textbf{a,b} Calculated pump--probe response (red area) for a 1500 nm probe, with forward (a) and backward (b) propagating THz pump. Theoretical spectra are compared with the FFT of the experimental signal (black curve) evaluated over the full time window (inset). FFTs of the incident or transmitted THz drive generated with a  1500 nm NIR pump are shown as gray areas. \textbf{c} Frequency dependence of the intrinsic bare-phonon damping rate $\gamma$ obtained from fitting experimental spectra with Eqs.\ \eqref{for_prop} and \eqref{back_prop}. The black curve represents the tanh-like function $\gamma(\Omega)$ discussed in the main text, with $\gamma_0=3.15\text{ THz}$, $\gamma_\infty=0.96\text{ THz}$, $\Omega_0/2\pi=2.81\text{ THz}$, and $\Delta/2\pi=0.36 \text{ THz}$. Dashed horizontal lines indicate $\gamma_0$ (dark gray) and $\gamma_\infty$ (light gray). The light-blue box indicates the region where further anharmonic couplings could affect the damping rate, as discussed in the main text. \textbf{d} Imaginary part of the phonon-polariton dispersion (black) computed from Eq.\ \eqref{wave_eq} using $\gamma(\Omega)$. Dashed curves represent the imaginary dispersion calculated with $\gamma_0$ (dark gray) and $\gamma_\infty$ (light gray). Dots represent the experimental phonon-polariton damping rates, as in Fig. \ref{expDiscussion}d. The gray shaded area is the experimental uncertainty of the phonon-polariton damping rate measured with ISRS in Ref.\ \cite{Knighton2018}. The light-blue box indicates the spectral region in which a second peak was observed in Ref.\ \cite{Knighton2018}, as discussed in the main text.}
    \label{theoryResults}
\end{figure*}

We determine $\text E^{(1)}_\text{NL}(\Omega)$ and $\text E^{(2)}_\text{NL}(\Omega)$ by numerically solving the integral in Eqs.\ \eqref{for_prop} and \eqref{back_prop},
using the experimental incident THz drive in $\text A_\text{p}^t$, and the measured transmitted THz pulse in $\text A_\text{p}^{r}$, while we model the probe field $\text A_\text{pr}^t$ as a Gaussian pulse of 100 fs-duration for 800 nm, and 120 fs for other wavelengths. Propagation of the pulses through the interfaces of the sample are taken into account with transmission and reflection coefficients, as discussed in Appendix \ref{theory}. The resulting frequency spectra are reported in Appendix \ref{comparison} and show a close match with the experimental FFT data over the entire range of probe wavelengths.
In Figs.\ \ref{theoryResults}a and \ref{theoryResults}b we show the calculated spectra for a 1500 nm probe as a red-shaded area, compared with the experimental data plotted as a solid black line. For reference, the incident and transmitted THz pump used in the calculations are shown as shaded gray areas. The insets display the time-domain window used for the FFT of the experimental data from Fig.\ \ref{expResults}(a), which includes the early signal induced by the THz drive. 
This early signal gives rise to the broad low-frequency shoulders observed in the spectra, in correspondence of the pump peak (gray dashed lines in Figs.\ \ref{expResults}b,c, and reported in Fig.\ \ref{theoryResults}a,b), and are well reproduced by the theoretical model.
Satellite features around 3 THz in Signal 2, arising from residual undamped oscillations of Signal 1 and indicated by the gray arrow in Fig.\ \ref{}c, are not reproduced by Eq.\ \eqref{back_prop}.

For the calculations, we take the optical refractive indices $n_o$ and $n_e$ from literature \cite{refractiveindex}. Values for $\omega_\text{TO}$, $\omega_\text{LO}$ and $\varepsilon_\infty$ are obtained from Ref.\ \cite{Knighton2018} while the damping parameter $\gamma$, entering in the THz refractive index in Eq.\ \eqref{nTHz} and in the bare-phonon propagator of Eq.\ \eqref{kernel}, is a free fitting parameter (Appendix \ref{comparison}). The extracted values for $\gamma$ are shown in Fig.\ \ref{theoryResults}c, as a function of the frequency of the main peak of the spectra, revealing a non-constant behavior. 
In particular, we find that $\gamma$ of the $E(\text{TO}_1)$ branch exhibits a smooth, step-like dependence on frequency over the experimentally accessible range. This range lies far from the TO resonance and therefore falls outside the domain accessible to conventional linear-response techniques, which probe the phonon damping only in the immediate vicinity of the resonance as dictated by the narrow refractive index Eq.\ \eqref{nTHz}.
We parameterize this step-like dependence with a phenomenological hyperbolic tangent $\gamma(\Omega)=\gamma_{\infty}+\frac{\gamma_0-\gamma_\infty}{2}[1-\tanh(\frac{\Omega-\Omega_0}{\Delta})]$ (black curve in Fig.\ \ref{theoryResults}c), where $\gamma_\infty$ and $\gamma_0$ represent high- and low-frequency damping parameters respectively, $\Omega_0$ is the threshold frequency at which the crossover between the two regimes occur, and $\Delta$ controls the width of the step-like behavior. The curve $\gamma(\Omega)$ is then used in Eq.\ \eqref{nTHz} to re-evaluate the polariton dispersion via Eq.\ \eqref{wave_eq}. While the real part of the dispersion is unaffected by the frequency-dependent damping rate, the imaginary part displays an anomalous behavior, as shown in Fig.\ \ref{theoryResults}d.
Qualitatively, this feature can be interpreted as a crossover between two regimes: at low frequencies, the polariton damping follows the dispersion obtained with a constant $\gamma_0=3.15\text{ THz}$, while at higher frequencies it approaches the curve corresponding to $\gamma_\infty=0.96\text{ THz}$ (gray dashed lines).

We note that the polariton damping rate, obtained as the imaginary part of the solution of Eq. \eqref{wave_eq} and extracted from the FWHM of the experimental traces, is the phonon damping modified by the coupling to light. By contrast, the bare-phonon damping $\gamma(\Omega)$ obtained through our fitting procedure is a less ambiguous quantity for understanding the interactions between the phonon and other modes.
\subsection*{Discussion}\label{sec3}
The theoretical model is able to reproduce experimental features that cannot be captured with simple Lorentzian fits of the main peaks of the spectra, such as the broad low-frequency shoulders, while also providing a microscopic picture of the three-wave-mixing mechanism. As described by Eqs.\ \eqref{for_prop} and \eqref{back_prop}, in TP-RP the THz pulse travels through the sample and then coherently excites the bare $E$ phonon modes, as expressed by the nonlinear interaction kernel $\text{K}^{(2)}(\omega^\prime,\Omega)$ in Eq.\ \eqref{kernel}. 
Because the polaritonic response enters the three-wave-mixing signal exclusively via propagation effects governed by the THz refractive index $n_\text{THz}(\Omega)$ inside the material, and is not affecting the kernel, the THz field itself is propagating as the phonon-polariton.
This behavior is fundamentally different from ISRS, where the THz refractive index does not affect the propagation of the pulses, and the polaritonic effects are instead encoded in the nonlinear interaction kernel that describes the internal response \cite{Dougherty1992,Sellati2025}.

By accounting for all contributions to the linewidth of the pump–probe signal within a unified framework, our theoretical formalism, combined with the enhanced experimental accuracy of the TP-RP approach, enables a reliable extraction of the bare-phonon damping parameter $\gamma$. 
This is essential to disentangle intrinsic phonon dissipation from extrinsic broadening effects arising from the coupling to light.
With this procedure, we can attribute the anomalous polariton damping rate reported in Refs.\ \cite{Wiederrecht1995,Schwarz1996,Crimmins2002,Knighton2018} to the frequency dependence of the intrinsic phonon damping $\gamma(\Omega)$,
allowing for a discussion
of its microscopic origin. 
In particular, the step-like behavior of $\gamma(\Omega)$ suggests that the phonon decays into a continuum of excitations below $\Omega_0 = 2.81$ THz, resulting in a high damping rate $\gamma_0$, while this decay channel is suppressed above this frequency, leading to a lower damping rate $\gamma_\infty$.
A natural candidate for this decay mechanism is the anharmonic coupling with acoustic phonons, whose dispersive branches reach frequencies between $2.5$ THz and $3$ THz at the Brillouin-zone boundaries \cite{Parlinski2000}, comparable to the value obtained for $\Omega_0$.
Additional calculations of the anharmonic scattering with acoustic phonons will be required in future works to validate this hypothesis. Experimentally, measurements in the low-frequency regime below 1 THz would be particularly relevant, as $\gamma(\Omega)$ is expected to follow the acoustic-phonon density of states in this range.

A second peak-like feature around $k\sim6000\text{ cm}^{-1}$, reported in Ref.\ \cite{Knighton2018} and highlighted by the light-blue box in Fig.\ \ref{theoryResults}d, suggests the presence of a similar anharmonic mechanism further affecting $\gamma(\Omega)$ in the frequency range between $3.2\text{ THz}$ and $3.8\text{ THz}$.
Accessing this regime, highlighted by a second light-blue box in Fig.\ \ref{theoryResults}c, requires TP-RP measurements with probe wavelengths between 800 nm and 1200 nm. Extending the measurements to this frequency range in future work would help cover the gap, and provide further information on the anomalous phonon-polariton damping rate and the underlying anharmonic decay processes in LiNbO$_3$.

\section*{Conclusion}\label{sec4}
In summary, we demonstrate that TP-RP offers a new and versatile way of mapping phonon-polariton dispersions in polar materials, shown here for LiNbO$_{3}$.
Using a broadband THz pump and a tunable NIR wavelength probe, we exploit the phase-matching of the three-wave mixing process to obtain the momentum resolution needed to reconstruct the dispersion. 
THz pumping enables access to the polaritonic dispersion through a mechanism fundamentally different from all-optical approaches \cite{Dougherty1992,Sellati2025}.
To provide an unambiguous microscopic interpretation of the experimental observations, we therefore develop a theoretical framework that goes beyond a Lorentz-oscillator description by combining many-body techniques with a systematic treatment of propagation effects, using only the intrinsic damping of the bare phonon as a free fitting parameter.

In particular, we demonstrate that a THz pump enables temporal separation of forward- and backward-propagating signals, and that analyzing the FFT in a window after the THz drive effectively isolates the polaritonic response. This procedure provides a clean strategy to reconstruct both the real and imaginary parts of the dispersion.
Finally, by combining experiment and theory, we extract the intrinsic phonon damping rate in LiNbO$_3$ and reveal its nontrivial frequency dependence, possibly arising from decay into a continuum of excitations.  
While this behavior could in principle be observed with four-wave-mixing approaches such as ISRS, it would require a substantially more involved theoretical framework and would hinder a reliable determination of the intrinsic damping.

The technique developed here can be applied more broadly to map polaritonic dispersions in materials with thicknesses ranging from a few tens of micrometers up to the limit set by absorption of the propagating pulses. It also allows selective excitation of specific polaritonic branches by tailoring the THz spectral shape. The method is, however, limited to non-centrosymmetric crystals and requires phase-matching conditions to be satisfied.
Our approach complements recently published methods for mapping polaritonic dispersion based on four-wave-mixing scatterings, whose optical pumping does not offer direct phonon excitation and is thus less efficient \cite{Knighton2018,Dastrup2017}, as well as second-harmonic-generation techniques, whose approach is limited to materials with a bandgap larger than the second-harmonic photon energy \cite{Luo2024}.
The unique temporal separation of signals offered by TP–RP opens new opportunities for advanced techniques, such as 2D THz spectroscopy, by clearly resolving overlapping contributions. 
In addition, the broadband THz pump can resonantly drive multiple modes across its spectral bandwidth through direct excitation, enabling the investigation of a broader range of couplings than those accessible with ISRS \cite{Blake2022,Johnson2019,Knighton2019}.

\section*{Methods}\label{sec5}

\noindent 
\textbf{THz pump, variable probe measurements} 

\noindent 
For pump and probe beams generation, we split the output of an optical parametric amplifier (OPA) driven by a Coherent Libra system (800 nm, 100 fs, 1 kHz repetition rate) into two beams using a 90:10 beam sampler.
We use the higher-energy transmission beam (90\%) to generate a modulated NIR pump, tuned between 1250 nm and 1550 nm and chopped at 500 Hz, and we send it on a 520 $\mu$m-thick PNPA crystal for the generation of broadband THz pulses \cite{Stillhart2008,Hauri2011,Rader2022}.
We use the lower-energy transmission beam (10\%), tuned to the same wavelength range, as probe pulses for our 
experiment. 

For the pump--probe measurements, we focus the emitted THz pump and the NIR probe in a collinear geometry onto a 500 $\mu\text{m}$-thick, x-cut [100] LiNbO$_3$ single crystal purchased from MTI Corporation, as described in Appendix \ref{setup}.
We use a mechanical delay stage to control the temporal delay between the THz pump and NIR probe pulses and track the ellipticity changes of the NIR probe using a balanced detection scheme consisting of a quarter-wave plate, 
a Wollaston prism, and balanced diodes.
We measure the pump--probe dynamics over a 30 $\text{ps}$ time window by scanning the delay stage in 5 $\mu\text{m}$ steps and acquiring 350 shots per step.  
We also carry out the pump--probe measurements using an 800 nm beam picked off from the Ti:sapphire laser output used to seed the OPA. In this case, we guide the 800 nm beam onto the sample through the same path used for the NIR probes and use a THz-generating beam fixed at 1450 nm, as schematized in Appendix \ref{setup}.

For a realistic simulation of the pump--probe response, we characterized the THz spectral profiles generated by each NIR pump both at the sample position and after transmission through the sample (Appendix \ref{field}), using an 800 nm gate beam. We show the PNPA-generated THz waveforms by the 1450 nm and 1500 nm NIR pump in Figs.\ \ref{}b,c and \ref{theoryResults}a,b, respectively, while the complete set is in Fig.\ \ref{thzFFTs}.

\noindent
\textbf{Numerical calculations and fitting of the data}

\noindent All numerical calculations and the fitting of experimental data were performed using Wolfram Mathematica. Experimental peak frequencies and linewidths were extracted using MATLAB.

%
{\scriptsize
\bibliography{sn-bibliography}}
%
%
\section*{Acknowledgements}
We thank Janine Zemp-Dössegger for her assistance with preliminary experiments, and Dominik Juraschek and Michael M. Fechner for their contributions to early-stage discussions of this work. We also thank Kimball Nielson for his 3D Modeling guidance with Fig. 1b. We thank Ravi Finn, Mikayla Hunt, and Aayla Wheatley for their participation in the data collection.
\section*{Authors contributions}
J. A. Johnson conceived the experiment and, together with R. Acampora, coordinated peer collaboration. R. Acampora and M. Biggs performed the experiments and processed the data. N. Sellati and M. Udina developed the theory and performed the simulations.  
L. Benfatto, E. Abreu, S. L. Johnson and J. A. Johnson supervised the research. 
R. Acampora, M. Biggs, and N. Sellati wrote the manuscript with contributions from all authors, and all the authors contributed to the discussion and interpretation of results.
\section*{Competing interests}
These authors declare no competing interests.
\section*{Additional Information}
\noindent
\textbf{Funding.}
J.A.J. and M.F.B. acknowledge financial support by Gordon and Betty Moore Foundation (grant DOI 10.37807/GBMF12958); 
N.S.\ and L.B.\ acknowledge financial support by the European Union under the project MORE-TEM ERC-SYN (Grant No.\ 951215); by Sapienza University under the project Ateneo (Grants No.\ RM123188E357C540, and No.\ 
RP124190A63FAA97); and by the Italian Ministry of Education, University and Research under Project PRIN2022-CoInEx (Grant No.\ 2022WS9MS4).
E.A.\ acknowledges support from Schweizerischer Nation-
alfonds zur Förderung der Wissenschaftlichen Forschung through Starting Grant No.\ TMSGI2\_211211.
S.L.J.\ acknowledges support from the Swiss National Science Foundation (project 200021-228373).
\textbf{Supplementary information.}
The supplementary information for this paper is included as an Appendix. 
\textbf{Ethics approval and consent to participate.}
Not applicable.
\textbf{Materials availability.}
The samples used in this study are available from the corresponding authors upon reasonable request.
\textbf{Code availability.}
The codes used in this study are available from the corresponding authors upon reasonable request.
%
\clearpage
\onecolumn
\begin{appendices}
\section{THz pump--birefringence probe setup}\label{setup}
\begin{figure}[h]
    \centering
    \includegraphics[width=0.8\linewidth]{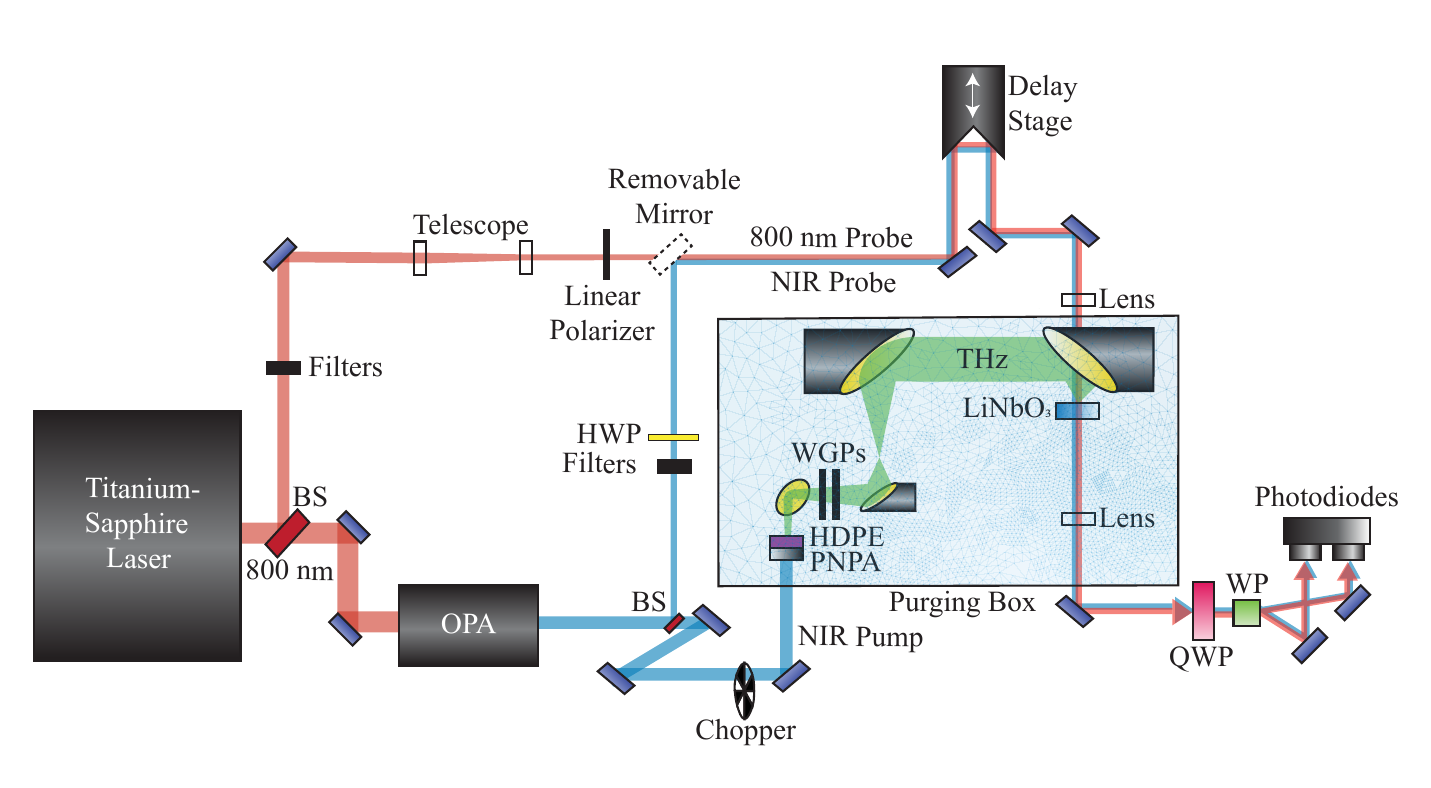}
    \caption{Schematic of the optical layout used for THz pump and near-infrared (NIR)/800 nm probe measurements. We split the output of a Coherent Libra Ti:sapphire amplifier into two beams using a 98:2 beam sampler. The higher-energy transmission beam (98\%) feeds a TOPAS Prime optical parametric amplifier (OPA) to generate tunable NIR pulses (1250–1550 nm), while the other one (2\%) provides an 800 nm probe beam. We generate THz pulses by optical rectification in a PNPA crystal and direct them to the sample using a sequence of three off-axis parabolic mirrors (1 inch $\diameter$, 1 inch EFL; 3 inches $\diameter$, 6 inches EFL; 3 inches $\diameter$, 2 inches EFL). We delay the probe beam, send it through a hole in the final parabolic mirror, and focus it onto the sample collinearly with the THz pump. We detect probe ellipticity changes using a balanced detection scheme consisting of a quarter-wave plate, a Wollaston prism, and balanced photodiodes. We place a lens after the sample to collimate the beam before sending it to the photodiodes. BS: beam sampler; OPA: optical parametric amplifier; PNPA: (E)-4-((4-nitrobenzylidene)-amino)-N-phenylaniline; HDPE: high-density polyethylene filter; WGPs: wire grid polarizers; QWP: quarter-wave plate; HWP: half-wave plate.}
    \label{expSetup}
\end{figure}
For the THz pump and birefringence probe measurements, we use the setup shown in Fig.\ \ref{expSetup} and a 500 $\mu\text{m}$-thick, x-cut [100] LiNbO$_3$ single crystal purchased from MTI Corporation.
The laser source is a Coherent Libra system delivering 800 nm pulses with 100 fs full-width-at-half-maximum (FWHM) at a repetition rate of 1 kHz. We split the high-power 800 nm output into two paths: one that directs the beam into a Light Conversion TOPAS Prime optical parametric amplifier (OPA) to generate tunable linearly polarized NIR pulses, with approximate FWHM 120 fs, and with wavelengths between 1250 nm and 1550 nm; and another path that bypasses the OPA. 

We direct the signal output from the OPA (depicted as blue in the figure) into a 90:10 beam sampler (BS) (10\% reflectance) and we send the transmitted NIR beam through an optical chopper to allow only every other pulse to pass without clipping, resulting in an effectively 500 Hz repetition rate. We measured the resulting beam energy to be approximately 0.2 to 0.7 mJ (depending on the NIR wavelength) and a 7.6 mm $1/e^2$ diameter at the 520 $\mu\text{m}$-thick PNPA crystal used for THz generation \cite{Stillhart2008,Hauri2011,Rader2022}. We place a 2.6 mm thick HDPE filter immediately after the THz-generation crystal to remove residual NIR beam, and we direct the THz light into a 2-inch gold mirror and a 3-parabolic-mirror focusing scheme consisting of a 1-inch diameter parabolic-mirror (effective focal length, EFL, 1 inch), a 3-inch diameter parabolic-mirror (EFL 6 inches), and a 3-inch diameter parabolic-mirror (EFL 2 inches). We find the resulting THz spot size at the sample position to be approximately 165 $\mu\text{m}$ $1/e^2$ spot size, measured with an electro-optic imaging method \cite{Hine2021,Lin2024,Jewell2025}.

For the probe pulses we use either the portion of the OPA beam reflected by the  BS or we send the remaining portion of the 800 nm beam from the amplifier (red path in the figure). We send the probe beam through a delay stage, then a central aperture in the third parabolic mirror, and finally we focus it onto the sample to a spot size of approximately 60 $\mu\text{m}$ measured with a camera in the IR range. We place neutral density filters before the sample to reduce the probe intensity and prevent sample damage. We control the temporal delay between the THz pump and NIR probe or 800 nm pulses with a mechanical delay stage. Lastly, we measure ellipticity changes of the NIR probe or 800 nm pulses using a balanced detection scheme consisting of a quarter-wave plate (QWP), with its extraordinary axis oriented at 45$^\circ$ relative to the incoming probe polarization, a Wollaston prism, set to split the beam into its vertical and horizontal polarization components, a lens to collimate the beam after the sample, and balanced Thorlabs IR photodiodes. 
\clearpage
\setcounter{equation}{0}
\setcounter{figure}{0}
%
\section{THz field strength characterization}\label{field}

We characterize the THz field both at the sample position and after transmission through the LiNbO$_3$ crystal with standard electro-optic (EO) sampling \cite{Planken2001}.

For the characterization of the THz field at the sample position, we use the setup depicted in Fig.\ \ref{expSetup} and replace the LiNbO$_3$ crystal with a $100~\mu\text{m}$-thick [110]-oriented GaP crystal bonded to a 1 mm-thick [100] GaP substrate. We take 2\% of the 800 nm beam from the Ti:sapphire laser output, attenuate it with filters, and use it to sample the THz transient. By controlling the temporal delay between the THz pump and the 800 nm gate pulses on the GaP crystal, we measure THz-induced transient ellipticity changes in the 800 nm gate beam polarization. For the measurements, we use a balanced detection scheme consisting of a quarter-wave plate (QWP), with its extraordinary axis oriented at 45$^\circ$ relative to the incoming gate polarization, a Wollaston prism, set to split the beam into its vertical and horizontal polarization components, and balanced photodiodes. 
We then obtain the temporal profile of the THz electric field by scanning the delay stage in the gate-beam path over a $30~\text{ps}$ time window in steps of $5~\mu\text{m}$ and acquiring 350 shots per step.
At high field strengths, the EO response becomes nonlinear, and the field estimate can be affected by saturation effects. We therefore use a pair of wire grid polarizers (WGPs, Fig.\ \ref{expSetup}), one to attenuate the THz beam and one to set the THz polarization perpendicular to the [001] axis of the GaP crystal, enabling field estimates at low transmission levels.

Once the temporal profile of the THz transient is obtained, we estimate the THz electric field strength ($\text{E}_{\mathrm{THz}}$) using the relationship between the THz-induced transient birefringence in the [110]-oriented GaP crystal and the resulting gate beam intensity variations, which are proportional to the applied field and can be calculated using \cite{Planken2001}:
\begin{align}\label{eqField}
\text{E}_{\mathrm{THz}} =
\frac{\lambda \sin^{-1}\!\left(\dfrac{I^{\text{on}}_1 - I^{\text{on}}_2}{I^{\text{off}}_1 + I^{\text{off}}_2}\right)}
{2 \pi L\, r_{41}\, n_0^{3}\, t_{\mathrm{GaP}}}.
\end{align}
Here, $I^{\text{off}}_1$ and $I^{\text{off}}_2$ are the positive intensities of the two orthogonal polarization components of the 800 nm gate beam in the absence of the THz field, while $I^{\text{on}}_1$ and $I^{\text{on}}_2$ are the corresponding intensities in the presence of the THz field. $L$ is the thickness of the GaP crystal, $n_0 = 3.2$ is the refractive index of GaP at the gate wavelength $\lambda=800$ nm \cite{Aspnes1983}, $r_{41} = 0.88~\mathrm{pm/V}$ is the EO coefficient of GaP \cite{Yu1989}, and $t_{\mathrm{GaP}} = 0.46$ is the transmission coefficient of GaP at the central frequency of the THz pulses \cite{Wu1997}.

In Table \ref{thz_peak_field}, we report the estimated peak-field strength values for THz generated by an NIR pump tuned between 1250 nm and 1550 nm. We estimated the peak-field strength values by taking the peak of the time-domain THz trace (defined as $\Delta I_{max} = I^{\text{on}}_{1,max} - I^{\text{on}}_{2,max}$) measured at 25\% transmission, and inserting it into Eq. (\ref{eqField}). The resulting values were then multiplied by 5 to obtain the corresponding 100\% field values.

For THz generated by a 1450 nm NIR pump, we also measured THz fields at transmission levels of 100\%, 41\%, 25\%, and 11\%, corresponding to polarizer angles of 0$^\circ$, 50$^\circ$, 60$^\circ$, and 70$^\circ$, respectively. Then, using the measurements acquired in the linear detection regime (i.e., 60$^\circ$ and 70$^\circ$), we estimated the uncertainty in the extrapolated peak field and obtained a standard deviation of 75 kV/cm. 

  \begin{table}[htbp]
\centering
\renewcommand{\arraystretch}{0.9}
\captionsetup{width=\textwidth}
        \centering
        \begin{tabular}{|c|c|}
            \hline
            \textbf{Rectified wavelength (nm)} & \textbf{THz Peak Field (kV/cm)}  \\  
            \hline
            1250 & 723 \\
            1300 & 746 \\
            1350 & 706 \\
            1400 & 701 \\
            1450 & 676 \\
            1500 & 572 \\
            1550 & 480 \\
            \hline
        \end{tabular}
        \vspace{10pt}
        \caption{Estimated THz peak electric field values estimated with Eq.~(\ref{eqField}) for each NIR wavelength used to pump the PNPA crystal. The values were obtained as described in the text. For pump--probe measurements using an 800 nm probe, we fixed the THz-generating beam at 1450 nm.}
        \label{thz_peak_field}
    \end{table}%
    \FloatBarrier

For the characterization of the THz pulse transmitted through the LiNbO$_3$ crystal, we collect the diverging THz radiation transmitted after the sample with a 3-inch-diameter off-axis parabolic mirror (EFL 2 inches), and focus it with a fifth parabolic mirror (2-inch-diameter, EFL 4 inches) onto a $100~\mu\text{m}$-thick [110]-oriented GaP crystal bonded to a 1 mm-thick [100] GaP substrate. 
We then measure the EO signal using the same balanced detection scheme and time window used for the characterization of the incident THz. As a reference, we also measure the THz signal transmitted through air at the detection position.

\begin{figure*}[h]
    \centering
        \centering
        \includegraphics[width=0.85\linewidth]{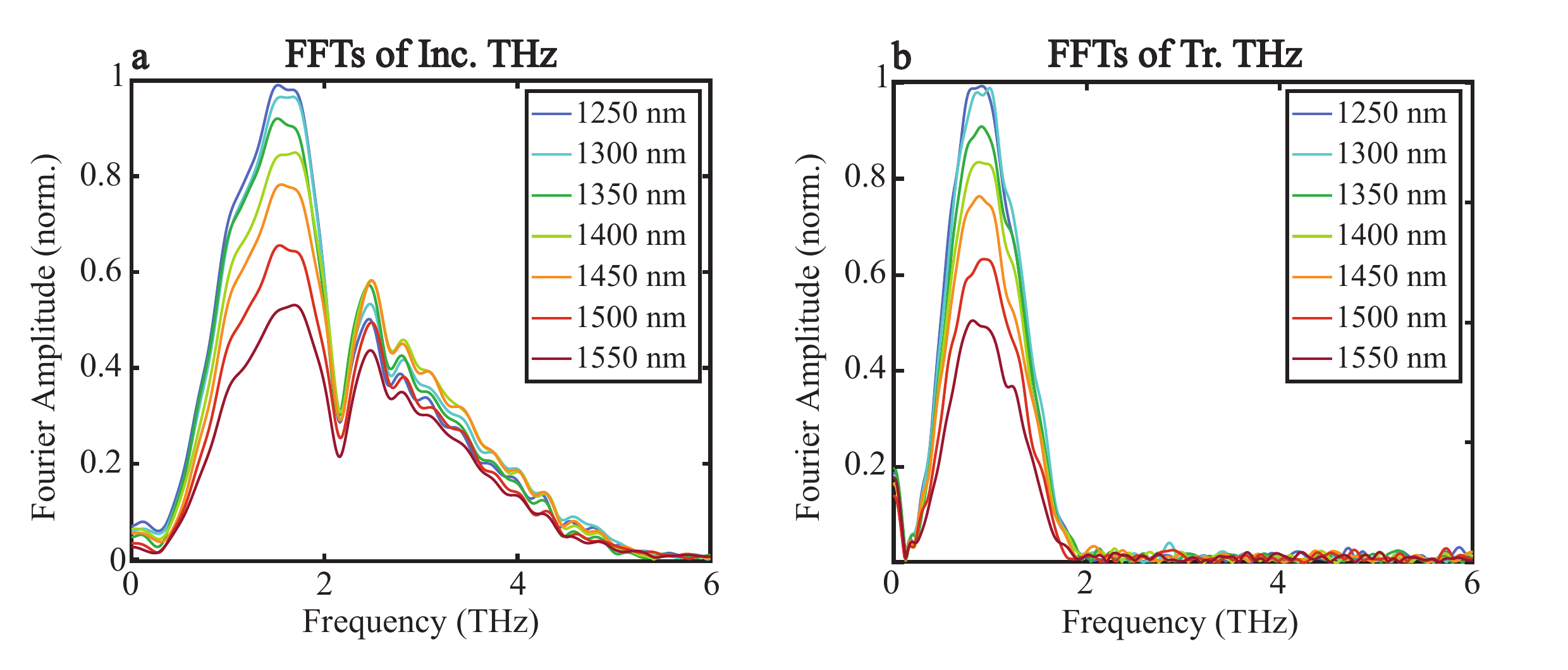}
        \caption{\textbf{a,b} Fourier spectra of THz traces generated via optical rectification of NIR pump wavelengths from 1250 to 1550 nm. The signals are normalized by the peak of the THz spectrum generated with a 1250 nm pump. (a) Incident THz spectra measured at the LiNbO$_3$ sample position with a GaP crystal. (b) Transmitted THz spectra recorded after propagation through the LiNbO$_3$ sample and detection with a GaP crystal.}
        \label{thzFFTs}
\end{figure*}

We show the FFTs of both the incident and transmitted THz fields for each NIR beam used to pump the PNPA crystal in Fig.\ \ref{thzFFTs}a and \ref{thzFFTs}b, respectively, normalized to the peak of the THz spectrum generated with a 1250 nm pump. The spectral content of the incident THz field ranges from 0.5 THz to 5.5 THz, and it is maximum at 1.5 THz, with a suppression of the spectral amplitude observed near 2 THz for all rectified NIR  wavelengths.
The transmitted THz spectra also exhibit a maximum near 1.5 THz, while the spectral components above 2 THz are strongly suppressed due to absorption within the LiNbO$_3$ sample.
For both incident and transmitted THz spectra, only small variations in spectral shape are observed across the different rectified pump wavelengths, indicating weak dependence of the optical rectification process on the NIR pump wavelength and suggesting only minor variation of the NIR group velocity within the PNPA crystal.

\begin{figure*}[h]
    \centering
        \centering
        \includegraphics[width=0.45\linewidth]{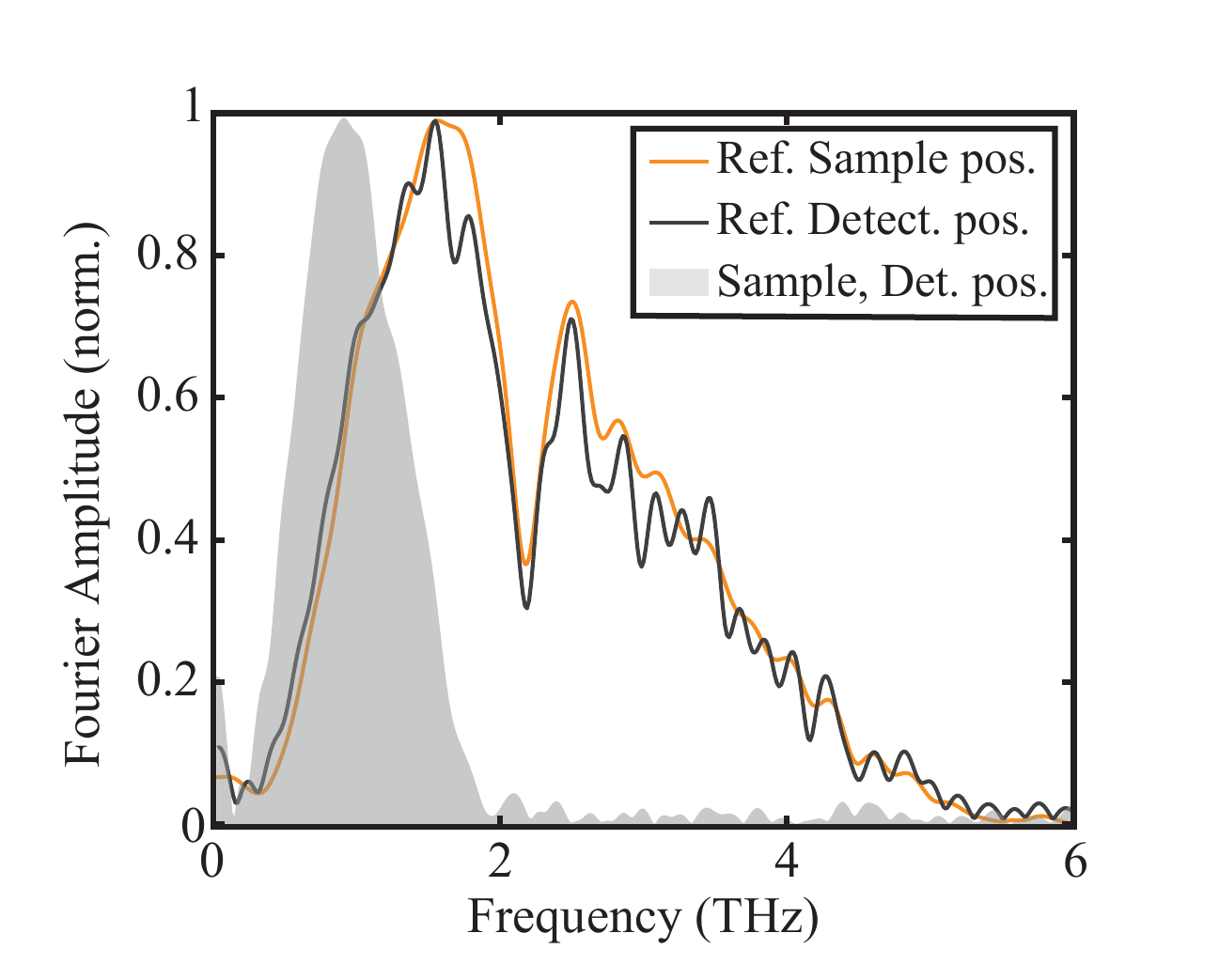}
        \caption{THz reference (Ref.) spectra generated with a 1450 nm pump and measured after propagation through air at the sample (orange) and detection (gray) positions using a GaP crystal and an 800 nm gate beam. For comparison, we also show the transmitted THz spectrum after propagation through the LiNbO$_3$ sample as a gray area.}
        \label{referencesTHz}
\end{figure*}

In Fig.~\ref{referencesTHz} we compare the THz spectrum transmitted through air and measured at the detection position with that measured at the sample position for THz pulses generated by optical rectification of a 1450 nm pump and detected using an 800 nm gate pulse. The two spectra exhibit comparable profiles, indicating that the refocusing process introduces negligible changes.

\clearpage
\setcounter{equation}{0}
\setcounter{figure}{0}
\setcounter{table}{0}

%
\section{Symmetry Selectivity of Pumping and Probing Schemes}\label{symmetry}

Pumping and probing processes in LiNbO$_3$ are constrained by symmetry considerations. 
Here we  neglect the spatial dependence of the fields, which is equivalent to assuming a thin crystal. This level of simplification is sufficient for our discussion of symmetry, and is removed in Appendix~\ref{phasematching} where we explicitly discuss phase matching.

LiNbO$_3$ crystallizes in the rhombohedral $R3c$ space group and belongs to the $C_{3v}$ point group. 
The vibrational modes decompose into three irreducible representations $A_1$, $A_2$, and $E$ \cite{Parlinski2000},
and the character table for this point group is shown in Table \ref{character_table}.
\begin{table}[h]
    \centering
    \renewcommand{\arraystretch}{1.2} 
    \begin{tabularx}{0.8\textwidth}{|c|c|c|c|>{\centering\arraybackslash}X|>{\centering\arraybackslash}X|}
        \hline
        & E & $2C_3(z)$ & $3\sigma_v$ & Linear, rotations & Quadratic \\
        \hline
        $A_1$ & 1 & 1 & 1 & z & $x^2+y^2, z^2$ \\
        \hline
        $A_2$ & 1 & 1 & -1 & $R_z$ & \\
        \hline
        $E$ & 2 & -1 & 0 & (x, y),\,($R_x$, $R_y$) & ($x^2-y^2$, xy),\,(xz, yz) \\
        \hline
    \end{tabularx}
    \caption{Character table for the C$_{3v}$ point group.}
    \label{character_table}
\end{table}\\
In the following we introduce the three-dimensional vector $\boldsymbol{Q}=(Q_1,Q_2,Q_3)$ in the space of the modes. Here, $Q_1$ describes the coordinate of the $E$-symmetry mode with dipole orientation along x, $Q_2$ the coordinate for the mode with dipole along y, and $Q_3$ the coordinate of the $A_1$ mode with dipole along z.

The linear coupling between a THz electric field $\textbf{E}(t)$ and infrared-active phonon modes is described by a contribution to the  potential energy density of the form $V_{lin}=\sum_i\text{E}_iP_i$, where  $P_i=\sum_\lambda{Z}^*_{i\lambda}Q_\lambda$ is the contribution to the polarization density that depends on the phonon mode coordinates. Here, the coupling constant ${Z}^*_{i\lambda}$ is the effective charge of the $\lambda$-th mode inducing a polarization along the $i$-th direction. The resulting classical equations of motion that describe the phonon driving read
\begin{align}\label{eq:C1}
    \ddot{Q}_\lambda(t) + 2\gamma_\lambda \dot{Q}_\lambda(t) + \omega_{\text{TO},\lambda}^2 Q_\lambda(t) = \sum_i \text{E}_i(t)Z^*_{i\lambda},
\end{align}
where $\omega_{\text{TO},\lambda}$ and $\gamma_\lambda$ refer to the TO frequency and damping rate of the $\lambda$-th mode, respectively.
Note that in our case \(Z_{i\lambda}^*\)
 is diagonal and so Eq.\ \ref{eq:C1} describes a system of three uncoupled equations.
 
Raman probing of the driven phonon involves an incoming and an outgoing optical field. The interaction is mediated by the phonon-induced modulation of the crystal polarizability, described by the Raman tensor $R_{ij}^{(\lambda)}$ of the $\lambda$-th mode.
The corresponding polarization density reads
\begin{align}
    P_i(t)=\sum_{j\lambda}R^{(\lambda)}_{ij} Q_\lambda(t) \text{E}_j(t).
\end{align}
%
%
The two-photon interaction is here considered instantaneous, i.e., the Raman tensor is taken as frequency independent. Raman activity of a phonon is determined by the transformation properties of the polarizability tensor, represented in the character table by the quadratic basis functions, $\text{x}^2+\text{y}^2$, $\text{z}^2$ for the $A_1$ mode, and ($\text{x}^2-\text{y}^2$, $\text{xy}$), ($\text{xz}$, $\text{yz}$) for the $E$ modes. The Raman tensors read explicitly \cite{Sanna2015}
\begin{align}\label{raman_tens}
    R^{(1)}=\begin{pmatrix}
        0 & -c & -d \\
        -c & 0 & 0 \\
        -d & 0 & 0
    \end{pmatrix},\qquad 
    R^{(2)}=\begin{pmatrix}
        c & 0 & 0 \\
        0 & -c & d \\
        0 & d & 0
    \end{pmatrix},\qquad
    R^{(3)}=\begin{pmatrix}
        a & 0 & 0 \\
        0 & a & 0 \\
        0 & 0 & b
    \end{pmatrix},
\end{align}

In the experimental conditions described in the main text, the THz and probe beam propagate in the x direction. Throughout the measurements, the THz is polarized along y, selectively driving the $E$-symmetry mode $Q_2$. The probe beam is polarized in the z direction, allowing us to probe the driven phonon through the scattered light polarized along y, as determined by the (3,2) component of $R^{(2)}$. 

\begin{figure}[h]
    \centering
    \includegraphics[width=0.35\linewidth]{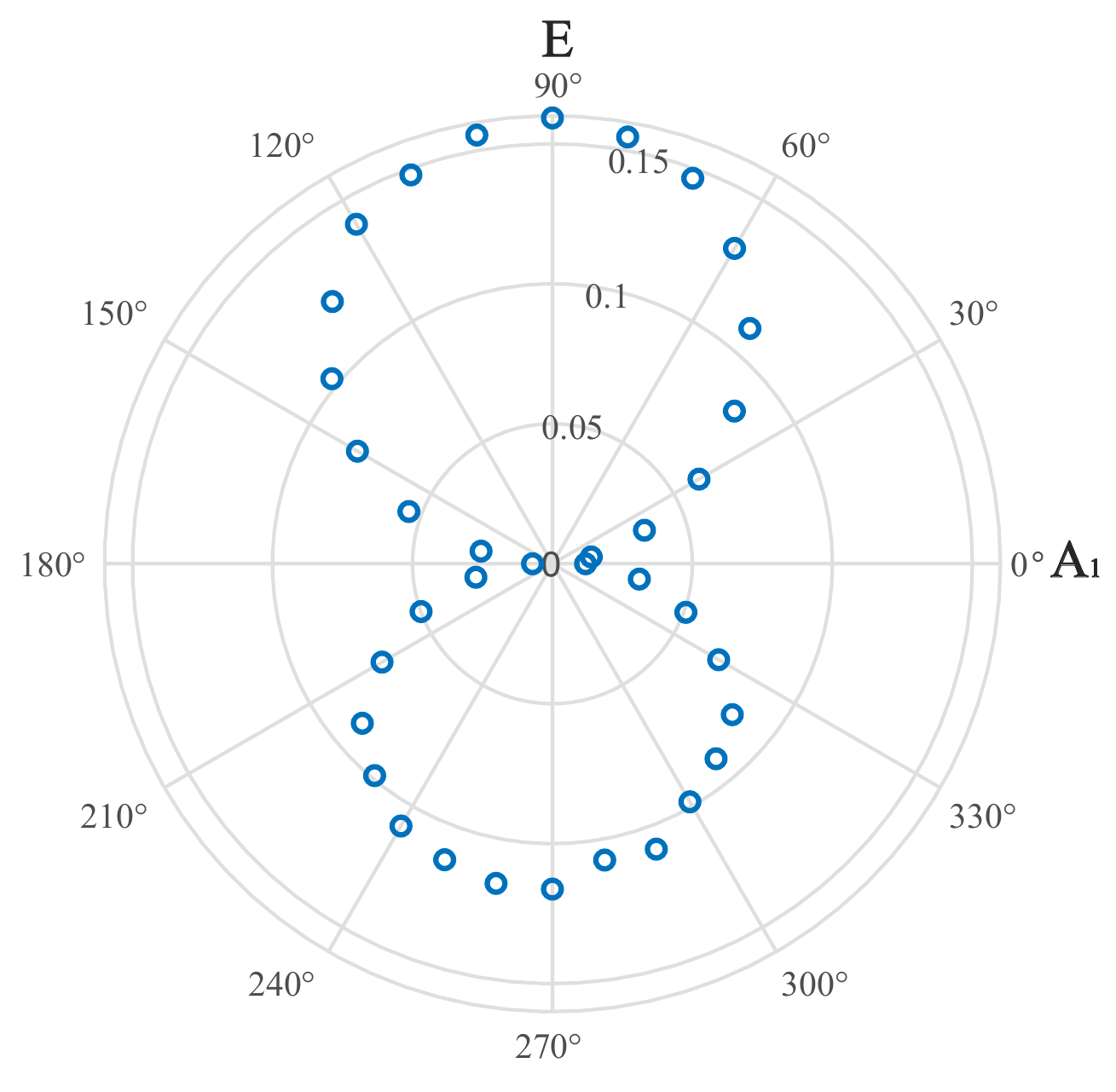}
    \caption{Angle-dependent peak-to-peak amplitude of the time-domain THz pump–800 nm probe signal measured as a function of sample rotation angle on a 275~$\mu\text{m}$-thick x-cut crystal. The THz pump was generated by optical rectification of 1450 nm pulses. During the measurement, we keep the THz pump polarization fixed while rotating the sample together with the probe polarization. An angle of $0^\circ$ corresponds to THz pump polarization aligned with the crystallographic z axis (nominally driving $A_1$ modes), while $90^\circ$ corresponds to THz pump polarization aligned with the y axis (driving $E$ modes). The observed  angular dependence has approximate two-fold symmetry and indicates selective excitation of $E$-symmetry phonons, with a maximum response when the pump is aligned with the crystallographic y axis.}
    \label{angDep}
\end{figure}

We perform angle-dependent pump–probe measurements on a 275~$\mu\text{m}$-thick x-cut crystal using an 800 nm probe and THz pulses generated by optical rectification of a 1450 nm beam. For these measurements, we rotate the sample together with two half-wave plates and record the pump–probe signal amplitude as a function of the sample rotation angle.
We rotate the first half-wave plate located after a linear polarizer and immediately before the sample as to probe always along the crystallographic z-axis, and we rotate a second half-wave plate, positioned immediately after the sample, to make sure that the outgoing probe polarization remains at 45$^\circ$ with respect to the extraordinary axis of the quarter-wave plate used in the balanced detection scheme described in Appendix \ref{setup}. Lastly, we keep the THz pump polarization fixed along the vertical direction in the laboratory reference frame.
We show the angular dependence of the response in Figure~\ref{angDep}. 
Here, $0^\circ$ corresponds to a THz pump polarized along the crystallographic z-axis, while $90^\circ$ corresponds to a THz pump polarized along the y-axis.
The response exhibits an approximately two-fold symmetric angular dependence. The deviation from exact two-fold symmetry is likely due to sample inhomogeneities and motion of the beam on the sample during the measurements.

\begin{figure}[h]
    \centering
    \includegraphics[width=0.7\linewidth]{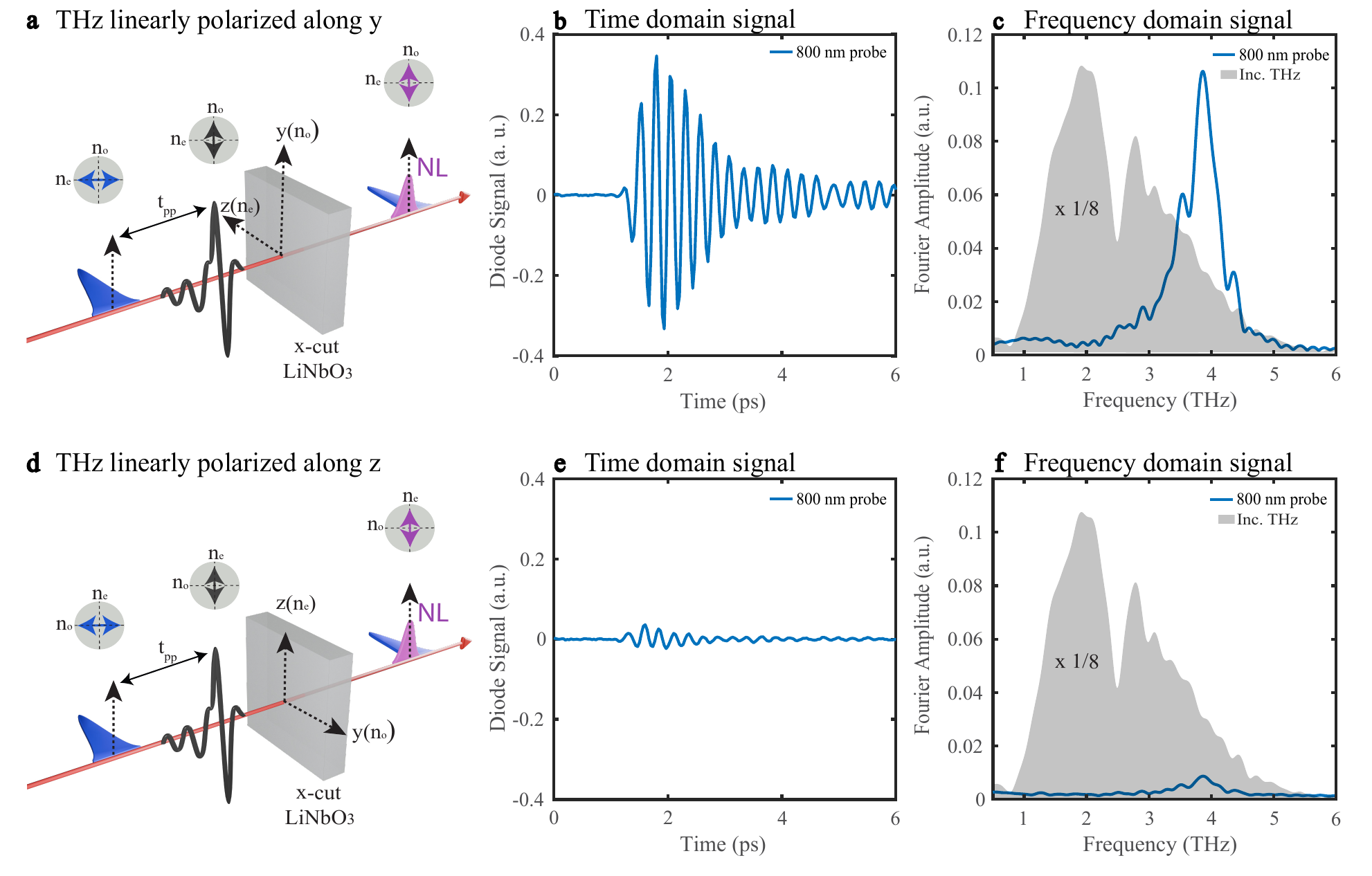}
    \caption{Comparison of the transient $\text{LiNbO}_3$ pump--probe responses for THz pump fields polarized along different crystallographic axes. \textbf{a,b,c} Configuration for the THz pump linearly polarized along the y axis: (a) Experimental schematic showing the x-cut crystal orientation; (b) time-domain transient oscillation trace; and (c) its Fourier transform spectrum (blue line) plotted alongside the spectrum of the incident THz pulse (gray area). \textbf{d,e,f} Configuration for the THz pump linearly polarized along the z axis: (d) Experimental schematic where the THz field is aligned to drive $A_1$ modes; (e) the resulting heavily suppressed time-domain signal; and (f) its corresponding Fourier transform spectrum.}
    \label{excitingEvsA1}
\end{figure}

To better illustrate the selectivity of the excitation process, we plot the pump--probe signals measured at $0^\circ$ and $90^\circ$  in Fig.\ \ref{excitingEvsA1}, both in the time and frequency domains. In the $0^\circ$ configuration, we still observe a weak but finite signal peaked at 3.8 THz. Since this frequency corresponds to the phase-matched $E$-mode for the 800 nm probe, we attribute this residual response to a slight misalignment of the THz polarization relative to the sample extraordinary axis, resulting in incomplete polarization extinction and unintended excitation of $E$-symmetry modes.

\clearpage
\setcounter{equation}{0}
\setcounter{figure}{0}
\setcounter{table}{0}

\section{Temporal separation between Signal 1 and Signal 2}\label{timesep}
To confirm that Signal 2 originates from the backward-propagating THz field, we show that the temporal separation between Signal 1 and Signal 2 in the pump–probe trace is consistent with the propagation time of the pump and probe pulses through the 500 $\mu$m sample.

\begin{figure}[h]
    \centering
    \hspace{20pt}
    \vspace{20pt}
    \includegraphics[width=0.4\linewidth]{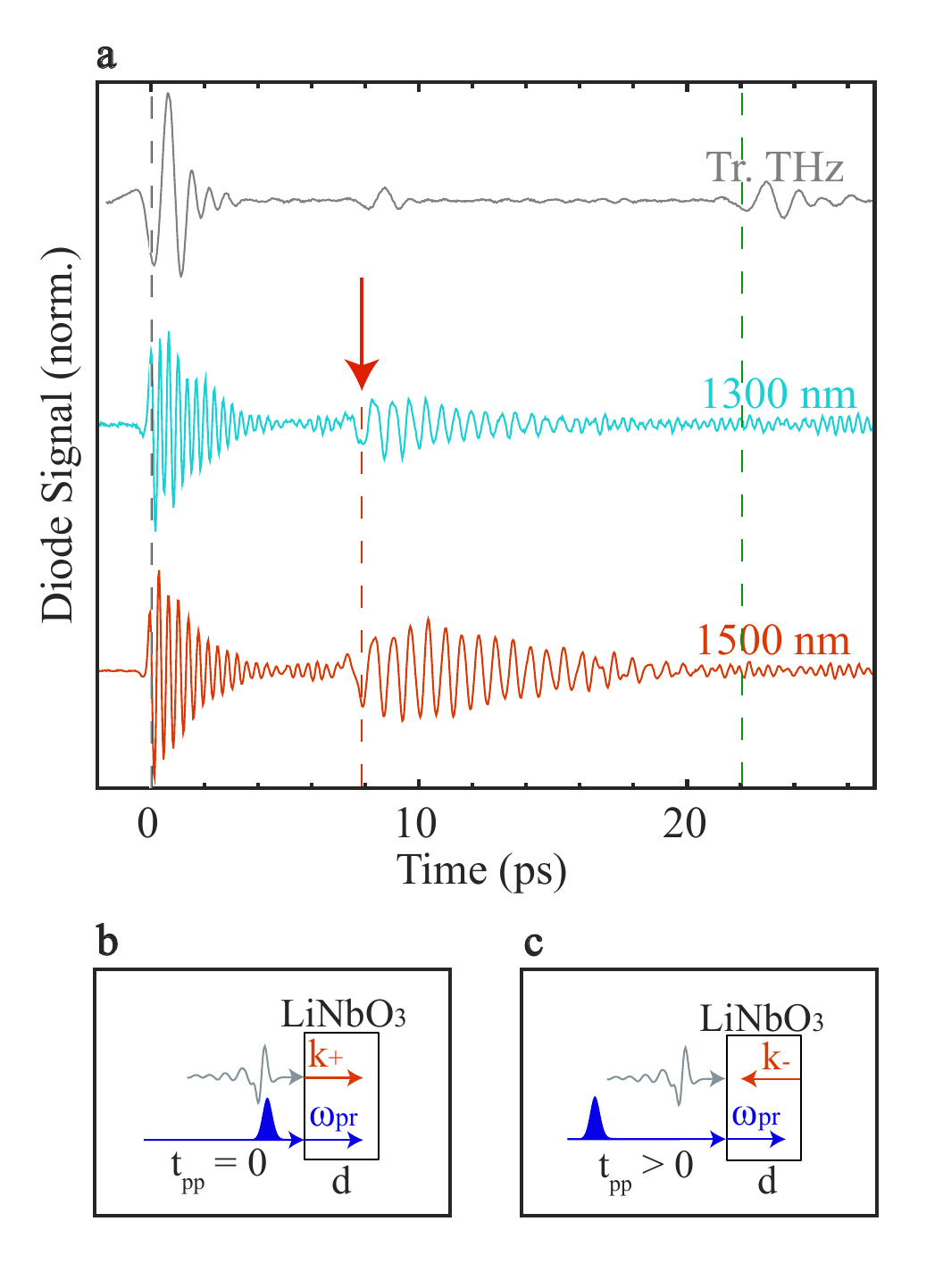}
    \caption{Time domain signals and reflection dynamics in a 500~$\mu$m-thick $\text{LiNbO}_3$ sample. \textbf{a} Normalized pump--probe time traces measured with 1300~nm (cyan) and 1500~nm (orange) probe wavelengths, compared alongside the transmitted THz field trace (grey). The vertical dashed grey line at $t = 0$~ps marks time zero for the three signals. The red arrow and dashed red line highlight the emergence of Signal 2 at $t_{pp}\approx 7.85$~ps, in agreement with the expected pump--probe delay $t_{pp}^{(\text{S}2)}$. The vertical dashed green line at $22$~ps highlights the round-trip internal reflection feature of the transmitted THz pulse, indicating a single-pass THz propagation time of $T_\text{pu} = 11$~ps. \textbf{b,c} Schematics illustrating the underlying physical mechanisms: (b) at $t_{pp} = 0$, the forward-propagating polariton ($k_+$) is launched by the incident THz field at the front interface; (c) at $t_{pp} > 0$, the backward-propagating polariton ($k_-$) is generated at the rear interface by the THz pulse reflecting off the back crystal boundary.}
    \label{signal2Timing}
\end{figure}

To extract the propagation time of the THz pulse, we identify an attenuated replica of the transmitted THz (Tr. THz) trace, appearing 22 ps after the main signal and corresponding to a component that has undergone a round trip through the sample (Fig.\ \ref{signal2Timing}a, vertical dashed green line). We thus infer that the time required for the THz light to pass once through the sample is $T_\text{pu}=11$ ps. We then compute the propagation time of the probe pulses,
\begin{align}
    T_\text{pr}=\frac{n_g}{c}d=3.75\text{ ps},
\end{align}
using a group refractive index $n_g=2.25$ for 1200 - 1500 nm probes along the extraordinary axis \cite{Nelson1974}, and a sample thickness $d=500\,\mu\text{m}$.

The time at which Signal 2 appears in the pump--probe time trace corresponds to the delay $t_{pp}$ for which pump and probe arrive simultaneously at the rear interface of the sample. This occurs at
\begin{align}
    t_{pp}^{(\text{S}2)}=T_\text{pu}-T_\text{pr}=7.25 \text{ ps}.
\end{align}
This estimate is in agreement with the experimental value of $t_{pp}$ at which Signal 2 emerges, indicated by a red arrow and dashed red line in Fig.\ \ref{signal2Timing}a, thus supporting the interpretation that Signal 2 originates from the backward-propagating THz pulse. In Fig.\ \ref{signal2Timing}b,c we report the geometries of the pulse interactions for forward- and backward-propagating signal, recalling Figs.\ \ref{expDiscussion}a,b of the main text.

\clearpage
\setcounter{equation}{0}
\setcounter{figure}{0}
\setcounter{table}{0}

\section{Complete pump--probe FFT spectra, time-domain windows, and experimental uncertainties}\label{ffts}

\begin{figure}[htbp]
    \hspace{-30pt}
    \includegraphics[width=1.2\textwidth]{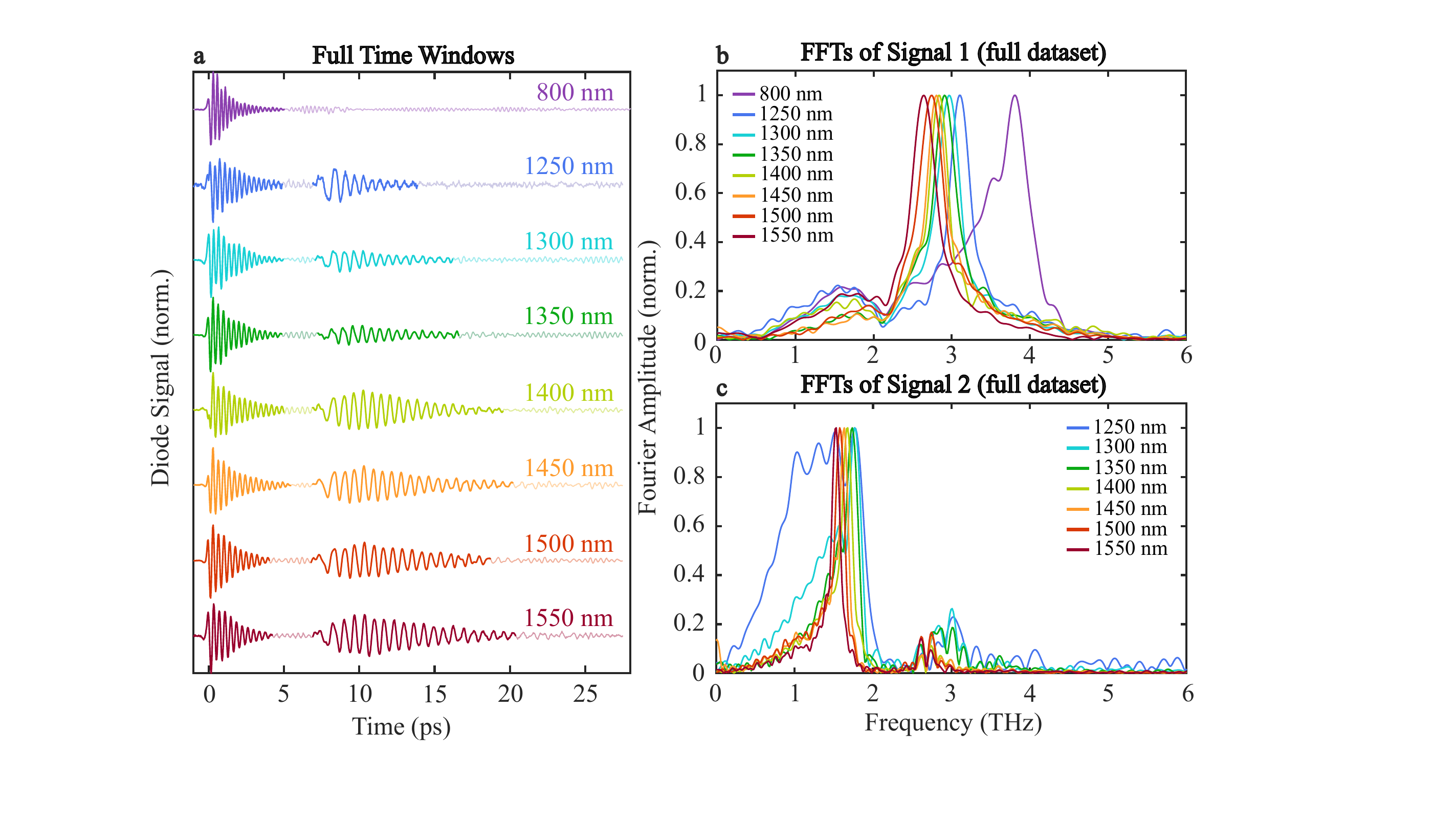}
    \caption{\textbf{a} Time-domain traces with the full FFT time windows highlighted. \textbf{b} Corresponding FFTs of Signal 1 calculated using the full time windows. \textbf{c} Corresponding FFTs of Signal 2 calculated using the full time windows.}
    \label{fullFFT}
\end{figure}

\begin{figure}[htbp]
    \hspace{-25pt}
    \includegraphics[width=1.2\textwidth]{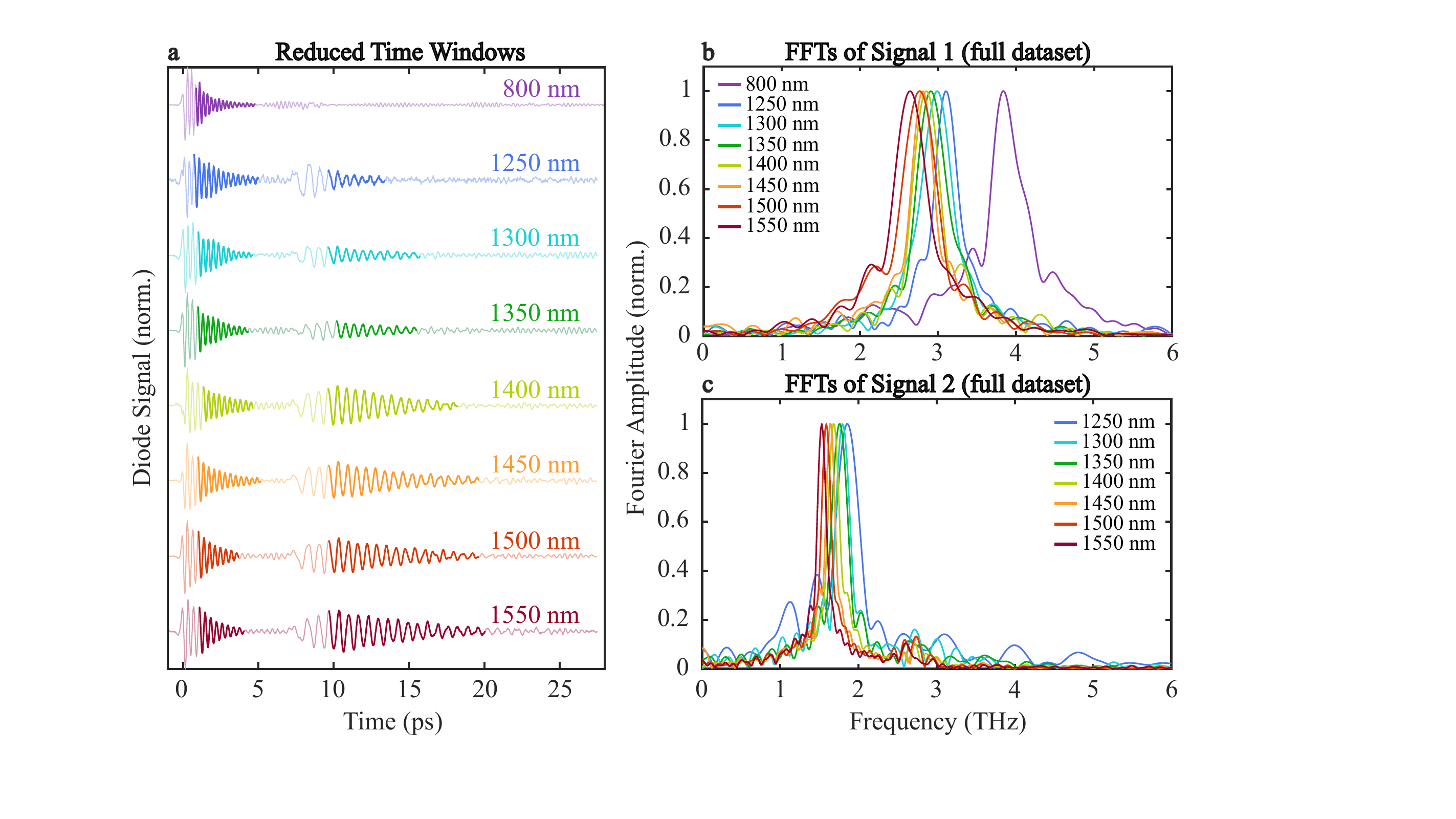}
    \caption{\textbf{a} Time traces with the reduced FFT time windows highlighted. \textbf{b} Corresponding FFTs of Signal 1 calculated using the reduced time windows. \textbf{c} Corresponding FFTs of Signal 2 calculated using the reduced time windows.}
    \label{reducedFFT}
\end{figure}

In Figs.~\ref{fullFFT} and \ref{reducedFFT}, we show the time evolution of the measured pump--probe signals together with their corresponding fast Fourier transforms (FFTs), reproducing Figs.~\ref{}a,b,c,d,e of the main text and including the FFTs for the remaining wavelengths (Figs.~\ref{fullFFT}b,c and \ref{reducedFFT}b,c). The dark-colored lines in the time traces indicate the full (Fig.\ \ref{fullFFT}) and reduced (Fig.\ \ref{reducedFFT}) time windows used to compute the FFTs of Signals 1 and 2, while we report the precise time intervals used for the FFT calculations in Tables~\ref{table_figE1} and \ref{table_figE2}.

\begin{table}[htbp]
\centering
\renewcommand{\arraystretch}{0.9}
\begin{tabularx}{\textwidth}{|X|X|X|X|X|X|X|}
\hline
\makecell{Probe \\ Wavelength \\ (nm)} &
\makecell{Sig. 1 start\\ window \\ (ps)} &
\makecell{Sig. 1 end\\ window \\ (ps)} &
\makecell{Total  \\ window \\ (ps)} &
\makecell{Sig. 2 start\\ window \\ (ps)} &
\makecell{Sig. 2 end \\ window \\ (ps)} &
\makecell{Total \\ window\\ (ps)} \\
\hline
800 &	-0.93 &	5.04 &	5.97 &	N/A &	N/A &	N/A\\
\hline
1250 &	-1.17 &	4.90 &	6.07 &	6.90 &	13.88 &	6.97\\
\hline
1300 &	-1.33 &	4.97 &	6.30 &	7.07 &	16.21 &	9.14\\
\hline
1350 &	-1.03 &	4.87 &	5.90 &	7.24 &	16.61 &	9.37\\
\hline
1400 &	-1.13 &	5.00 &	6.14 &	6.87 &	19.55 &	12.68\\
\hline
1450 &	-0.97 &	5.47 &	6.44 &	6.87 &	20.21 &	13.34\\
\hline
1500 &	-1.07 &	4.04 &	5.10 &	6.77 &	18.71 &	11.94\\
\hline
1550 &	-1.03 &	4.24 &	5.27 &	6.90 &	20.38 &	13.48\\
\hline
\end{tabularx}
\caption{Start and end times of the full time windows used for the Fourier transforms of Signal 1 and Signal 2.}
\label{table_figE1}
\end{table}
\newpage

\begin{table}[htbp]
\centering
\renewcommand{\arraystretch}{0.9}
\begin{tabularx}{\textwidth}{|X|X|X|X|X|X|X|}
\hline
\makecell{Probe \\ Wavelength \\ (nm)} &
\makecell{Sig. 1 start\\ window \\ (ps)} &
\makecell{Sig. 1 end\\ window \\ (ps)} &
\makecell{Total  \\ window \\ (ps)} &
\makecell{Sig. 2 start\\ window \\ (ps)} &
\makecell{Sig. 2 end \\ window \\ (ps)} &
\makecell{Total \\ window\\ (ps)} \\
\hline
800	& 0.87 &	4.81 &	3.94 &	N/A &	N/A &	N/A\\
\hline
1250 &	0.73 &	5.04 &	4.30 &	9.64 &	13.41 &	3.77\\
\hline
1300 & 	1.00 &	4.64 &	3.64 &	9.64 &	15.74 &	6.10\\
\hline
1350 &	1.00 &	4.34 &	3.34 &	10.17 &	15.51 &	5.34\\
\hline
1400 &	1.33 &	4.64 &	3.30 &	9.64 &	18.25 &	8.61\\
\hline
1450 &	1.00 &	5.17 &	4.17 &	9.64 &	19.65 &	10.01\\
\hline
1500 &	1.00 &	3.70 &	2.70 &	9.67 &	19.65 &	9.97\\
\hline
1550 &	1.07 &	4.07 &	3.00 &	9.67 &	20.05 &	10.37\\
\hline

\end{tabularx}
\caption{Start and end times of the reduced time windows used for the Fourier transforms of Signal 1 and Signal 2.}
\label{table_figE2}
\end{table}
\begin{table}[htbp]
\centering
\renewcommand{\arraystretch}{0.9}
\begin{tabularx}{\textwidth}{|X|X|X|X|X|X|X|}
\hline
\makecell{Probe \\ Wavelength \\ (nm)} &
\makecell{Average \\ Wavevector \\ (cm$^{-1}$)} &
\makecell{Std of  \\ Wavevector \\ (cm$^{-1}$)} &
\makecell{Average \\ Peak Freq. \\ (THz)} &
\makecell{Std of \\ Peak Freq. \\ (THz)} &
\makecell{Average \\ FWHM \\ (THz)} &
\makecell{Std of \\ FWHM \\ (THz)} \\
\hline
800 &	8072 &	11 &	3.82 &	0.02 &	0.65 &	0.01\\
\hline
1250 &	5231 &	10 &	3.12 &	0.02 &	0.40 &	0.07\\
\hline
1300 &	5022 &	21 &	3.01 &	0.04 &	0.41 &	0.04\\
\hline
1350 &	4832 &	21 &	2.92 &	0.04 &	0.36 &	0.00\\
\hline
1400 &	4682 &	31 &	2.90 &	0.07 &	0.51 &	0.32\\
\hline
1450 &	4595 &	14 &	2.81 &	0.03 &	0.36 &	0.04\\
\hline
1500 &	4362 &	10 &	2.74 &	0.02 &	0.41 &	0.02\\
\hline
1550 &	4215 &	10 &	2.67 &	0.02 &	0.41 &	0.02\\
\hline
\end{tabularx}
\caption{Experimental peak frequencies, FWHM, and wavevectors for Signal 1 at all probe wavelengths, taking the Fourier transform over the full time window. Values are averaged from 3 datasets.}
\label{full_signal1}
\end{table}
\begin{table}[htbp]
\centering
\renewcommand{\arraystretch}{0.9}
\begin{tabularx}{\textwidth}{|X|X|X|X|X|X|X|}
\hline
\makecell{Probe \\ Wavelength \\ (nm)} &
\makecell{Average \\ Wavevector \\ (cm$^{-1}$)} &
\makecell{Std of  \\ Wavevector \\ (cm$^{-1}$)} &
\makecell{Average \\ Peak Freq. \\ (THz)} &
\makecell{Std of \\ Peak Freq. \\ (THz)} &
\makecell{Average \\ FWHM \\ (THz)} &
\makecell{Std of \\ FWHM \\ (THz)} \\
\hline
800	& 8074	& 11 &	3.82 &	0.02	& 0.55	& 0.03\\
\hline
1250	& 5229	& 7	& 3.11	& 0.01	& 0.45 &	0.07\\
\hline
1300	& 5022 &	16 &	3.01	& 0.03	& 0.57	& 0.08\\
\hline
1350	& 4832	& 14	& 2.92	& 0.03 & 	0.48	& 0.01\\
\hline
1400	& 4673	& 14 &	2.88 &	0.03 &	0.41	& 0.04\\
\hline
1450	& 4598 &	12	& 2.82	& 0.03	& 0.38	& 0.03\\
\hline
1500	& 4369	& 10	& 2.76	& 0.02	& 0.54	& 0.05\\
\hline
1550	& 4222	& 16	& 2.68	& 0.03	& 0.54	& 0.03\\
\hline
\end{tabularx}
\caption{Experimental peak frequencies, FWHM, and wavevectors for Signal 1 at all probe wavelengths, taking the Fourier transform over the reduced time window. Values are averaged from 3 datasets.}
\label{reduced_signal1}
\end{table}
\begin{table}[htbp]
\centering
\renewcommand{\arraystretch}{0.9}
\begin{tabularx}{\textwidth}{|X|X|X|X|X|X|X|}
\hline
\makecell{Probe \\ Wavelength \\ (nm)} &
\makecell{Average \\ Wavevector \\ (cm$^{-1}$)} &
\makecell{Std of  \\ Wavevector \\ (cm$^{-1}$)} &
\makecell{Average \\ Peak Freq. \\ (THz)} &
\makecell{Std of \\ Peak Freq. \\ (THz)} &
\makecell{Average \\ FWHM \\ (THz)} &
\makecell{Std of \\ FWHM \\ (THz)} \\
\hline
1250 &	3147 &	171 &	1.36 &	0.37 &	1.08 &	0.05\\
\hline
1300 &	2922 &	106 &	1.50 &	0.23 &	1.00 &	0.45\\
\hline
1350 &	2670 &	14 &	1.74 &	0.03 &	0.56 &	0.47\\
\hline
1400 &	2549 &	12 &	1.69 &	0.03 &	0.21 &	0.05\\
\hline
1450 &	2529 &	10 &	1.64 &	0.02 &	0.17 &	0.05\\
\hline
1500 &	2359 &	4 &	1.58 &	0.01 &	0.14 &	0.02\\
\hline
1550 &	2266 &	8 &	1.54 &	0.02 &	0.14 &	0.01\\
\hline
\end{tabularx}
\caption{Experimental peak frequencies, FWHM, and wavevectors for Signal 2 at all probe wavelengths, taking the Fourier transform over the full time window. Values are averaged from 3 datasets.}
\label{full_signal2}
\end{table}
\begin{table}[htbp]
\centering
\renewcommand{\arraystretch}{0.9}
\begin{tabularx}{\textwidth}{|X|X|X|X|X|X|X|}
\hline
\makecell{Probe \\ Wavelength \\ (nm)} &
\makecell{Average \\ Wavevector \\ (cm$^{-1}$)} &
\makecell{Std of  \\ Wavevector \\ (cm$^{-1}$)} &
\makecell{Average \\ Peak Freq. \\ (THz)} &
\makecell{Std of \\ Peak Freq. \\ (THz)} &
\makecell{Average \\ FWHM \\ (THz)} &
\makecell{Std of \\ FWHM \\ (THz)} \\
\hline
1250 &	2893	& 22 &	1.91 &	0.05 & 0.61 &	0.26\\
\hline
1300 &	2772 &	14 &	1.82 &	0.03 &	0.43 &	0.27\\
\hline
1350 &	2657 &	14 &	1.76 &	0.03 &	0.31 &	0.10\\
\hline
1400 &	2543 &	14 &	1.71 &	0.03 &	0.21 &	0.03\\
\hline
1450 &	2520 &	14 &	1.66 &	0.03 &	0.24 &	0.08\\
\hline
1500 &	2353 &	4 &	1.59 &	0.01 &	0.15 &	0.02\\
\hline
1550 &	2264 &	7 &	1.55 &	0.01 &	0.17 &	0.02\\
\hline
\end{tabularx}
\caption{Experimental peak frequencies, FWHM, and wavevectors for Signal 2 at all probe wavelengths, taking the Fourier transform over the reduced time window. Values are averaged from 3 datasets.}
\label{reduced_signal2}
\end{table}

In Tables~\ref{full_signal1}, \ref{reduced_signal1}, \ref{full_signal2}, and \ref{reduced_signal2}, we report the relevant numerical values (i.e., Avg. Wavevector, Avg. Peak Freq., and Avg. FWHM) obtained by averaging over three independent datasets, together with the corresponding experimental uncertainties, calculated as the standard deviation (Std) of these datasets for each probe wavelength.

Specifically, Tables~\ref{full_signal1} and \ref{full_signal2} list the average frequencies, wavevectors, and FWHMs of the $k_+$ and $k_-$ signals obtained using the full FFT time windows reported in Table~\ref{table_figE1}. Similarly, Tables~\ref{reduced_signal1} and \ref{reduced_signal2} report the corresponding quantities obtained using the reduced FFT time windows reported in Table~\ref{table_figE2}. In both cases, we extract the FWHMs by measuring the widths of the FFT peaks at half of their maximum amplitude.

As discussed in the main text, Eqs.~(\ref{phase_matching1}) and (\ref{phase_matching2}) are valid only for a narrowband probe field, such as the one employed in our experiment ($\pm 10$ nm), for which the resulting spread in phase-matched frequencies remains small. To quantify this effect, we estimated the phase-matching spread for each probe wavelength by evaluating the phase-matching conditions at the lower and upper limits of the probe bandwidth. The resulting uncertainties in $k_+$ and $k_-$ are reported in Table~\ref{pm_spread}.

\begin{table}[htbp]
\centering
\renewcommand{\arraystretch}{0.8}
\begin{tabularx}{\textwidth}{|X|X|X|}
\hline
\makecell{Probe Wavelength (nm)} &
\makecell{Spread $k_-$ (cm$^{-1}$)} &
\makecell{Spread $k_+$ (cm$^{-1}$)} \\
\hline
800 &	72 &	93\\
\hline
1250 &	51 &	67\\
\hline
1300 &	55 &	72\\
\hline
1350 &	59 &	77\\
\hline
1400 &	54 &	71\\
\hline
1450 &	58 &	76\\
\hline
1500 & 54 & 70\\
\hline
1550 &	53 &	70\\
\hline
\end{tabularx}
\caption{Estimated spread of the phase-matched momenta for the different probe wavelengths.}
\label{pm_spread}
\end{table}

\clearpage
\setcounter{equation}{0}
\setcounter{figure}{0}
\setcounter{table}{0}

\section{Phase-matching equations for the $A_1$ mode}\label{phasematching}
As discussed in Appendix \ref{symmetry}, the Raman tensor of $A_1$-symmetry modes of point group $C_{3v}$ is diagonal, so that light scattering on them does not change its polarization. Because of this, phase-matching with the lowest $A_1$-symmetry phonon-polariton of LiNbO$_3$ cannot be realized, and its dispersion cannot be probed with TP-RP. To see this, we can explicitly rewrite the phase-matching conditions $k_{\pm}-k_\text{in}-k_\text{out}=0$ for this mode,
\begin{align}\label{phase-matchingA1}
    k_{+}=&\frac{n_o}{c}\Omega_+,\nonumber\\
    k_{-}=&\frac{n_o}{c}\Omega_-,
\end{align}
where we assumed the probe polarized along the ordinary axis. Because the probe polarization is unchanged in the nonlinear process, forward- and backward-going polaritons respect the same phase-matching condition, analogously to the case of a cubic material \cite{Sellati2025}.
To fulfill phase-matching, the momenta in Eqs.\ \eqref{phase-matchingA1} must match for at least one frequency $\Omega$ obtained from the wave equation 

\begin{align}\label{wave_eqA1}
    k=\frac{n_\text{THz}(\Omega)}{c}\Omega.
\end{align}

Here $n_\text{THz}(\Omega)$, Eq.\ \eqref{nTHz} of the main text, refers to the THz refractive index in the z direction, modulated by the $A_1(\text{TO}_1)$ phonon. For LiNbO$_3$, $\omega_\text{TO}/2\pi=7.55\text{ THz}$, $\omega_\text{LO}/2\pi=8.27\text{ THz}$ and $\varepsilon_{\infty}=16.5$ \cite{Barker1967}. The minimum value of the THz refractive index for $\Omega<\omega_\text{TO}$ is in this case $n_\text{THz}(0)=\sqrt{\varepsilon_\infty\omega_\text{LO}^2/\omega_\text{TO}^2}=4.45$. Because $n_o<n_\text{THz}(0)$ for visible and NIR probes \cite{refractiveindex}, the polariton momentum obtained from Eq.\ \eqref{wave_eqA1} is always bigger than the probe momentum given by Eq.\ \eqref{phase-matchingA1}, and phase-matching is never realized.
The same conclusion can be drawn for a probe polarized along the extraordinary axis, since $n_e<n_o$.

We note that this result is not general for $A_1$ phonon-polaritons. For a material with $n_o>n_\text{THz}(0)$, phase-matching can be realized for the frequency $\Omega$ such that $n_o=n_\text{THz}(\Omega)$, and the TP-RP technique can be applied to measure the dispersion curve. 
\clearpage
\setcounter{equation}{0}
\setcounter{figure}{0}
\setcounter{table}{0}

\section{Theory Formalism}\label{theory}
We here build on the theoretical formalism developed in Ref.\ \cite{Sellati2025} for cubic non-centrosymmetric crystals, and generalize it to uniaxial crystals. Considering the experimental setup described in Appendix \ref{setup}, we take the wavevectors of the external pulses along the x crystallographic axis. Because of this choice, the lattice oscillations are not coupled and we can treat the $A_1$- and $E$-symmetry modes separately. The general strategy involves computing the nonlinear second-order current through a many-body formalism, and then link it to the nonlinear signal measured outside the sample with a Maxwell-Fresnel approach for nonlinear processes, which takes into account all propagation effects of the pulses.
\subsection{Internal response}
To compute the internal response of the system, we employ a many-body effective-action formalism, in which the system is described by the partition function $\mathcal{Z}=\int\mathcal{D}[Q,\text{A}_\text{p},\text{A}_{pr}]e^{-S[Q,\text{A}_\text{p},\text{A}_\text{pr}]}$. Here, $S[Q,\text{A}_\text{p},\text{A}_\text{pr}]$ is the total action of the bare phonon $Q$, the pump field $\text{A}_\text{p}$, and the probe field $\text{A}_\text{pr}$. In this framework, the response is described by the second-order nonlinear current, which can be computed as the functional derivative of the effective action of the electromagnetic field $S_\text{eff}[\text{A}_p,\text{A}_\text{pr}]$ with respect to the scattered outgoing probe pulse,
\begin{align}\label{nl_curr}
    \text{J}^{(2)}(\omega,k)=-\frac{\partial S_\text{eff}[\text{A}_p,\text{A}_\text{pr}]}{\partial \text{A}_\text{pr}(-\omega,-k)/c}.
\end{align}
The effective action is obtained after integration in $\mathcal{Z}$ of all the internal degrees of freedom, represented in this case by the phonon mode $Q$.

The bare $E$-symmetry mode with dipole orientation along the crystallographic y direction $Q_2$ is described by the Gaussian action
\begin{align}\label{bare_ph_act}
    S_0[Q_2]=\sum_qD^{-1}(i\Omega_m)\big|Q_2(q)\big|^2,
\end{align}
with $q=(i\Omega_m,k)$ the four-momentum, $\Omega_m=2\pi m/T$ the bosonic Matsubara frequencies, and $T$ the temperature. The bare-phonon propagator reads, in Matsubara space,
\begin{align}\label{bare_ph}
    D(i\Omega_m)=-\frac{2}{\Omega_m^2+\omega_\text{TO}^2}.
\end{align}
We then introduce the coupling to an external perturbation, represented by the electromagnetic gauge potential. The THz pump field $\text{A}_{\text{p}}$ polarized along the y axis couples linearly to the phonon, as described by
\begin{align}\label{lin_coup}
S_\text{THz}[Q_2,\text{A}_\text{p}]=\sum_q\frac{\Omega_m}{c}Z^*\text{A}_{\text{p,y}}(q)Q_2(-q).
\end{align}
The linear coupling is mediated by the mode effective charge $Z^*$ associated with the $E(\text{TO}_1)$ mode. On the other hand, the Raman probe $\text{A}_{\text{pr}}$ interacts quadratically with the phonon:
\begin{align}\label{raman_coup}
S_\text{Raman}[Q_2,\text{A}_\text{pr}]=\sum_q\sum_{q^\prime}\sum_{q^{\prime\prime}}\frac{\Omega_m^\prime\Omega_m^{\prime\prime}}{c^2}Q_2(q)R\text{A}_{\text{pr,z}}(q^\prime)\text{A}_{\text{pr,y}}(q^{\prime\prime})\delta_{q^{\prime\prime},-q-q^\prime}.
\end{align}
Here, $R$ represents the phonon Raman tensor component of $R^{(2)}$, in Eq.\ \eqref{raman_tens}, that rotates the probe polarization from the extraordinary (z) to the ordinary (y) axis, i.e., the (3,2) component. In the following we suppress the spatial indices of the fields for compactness. The overall action that describes the system is then $S[Q_2,\text{A}_\text{p},\text{A}_\text{pr}]=S_0[Q_2]+S_\text{THz}[Q_2,\text{A}_\text{p}]+S_\text{Raman}[Q_2,\text{A}_\text{pr}]$. We note that the linear coupling Eq.\ \eqref{lin_coup} and the quadratic coupling Eq.\ \eqref{raman_coup} are simultaneously allowed only for a phonon that is both IR- and Raman-active.

By integration of the phonon field $Q_2$, we obtain all processes coming from $(\text{A}_\text{p}+\text{A}_\text{pr}^2)^2$. Since we are interested in describing three-wave-mixing processes, we retain only the term of first order in the THz field and of second order in the optical field, resulting in the effective action
\begin{align}
    S_\text{eff}[\text{A}_\text{p},\text{A}_\text{pr}]=-\frac{1}{2c^3}\sum_{q}\sum_{q^\prime}\sum_{q^{\prime\prime}}\Omega_m\Omega_m^\prime\Omega_m^{\prime\prime}\text{A}_\text{p}(q)\big[Z^*R\,D(i\Omega_m)\big]\text{A}_\text{pr}(q^\prime)\text{A}_\text{pr}(q^{\prime\prime})\delta_{q^{\prime\prime},-q-q^\prime}.
\end{align}
This action describes the second-order process in which a bare phonon is excited directly by the THz field, and then subsequently scatters with an optical photon. We can then compute the nonlinear current generated inside the material through Eq.\ \eqref{nl_curr}, yielding
\begin{align}\label{nl_curr2}
    \text{J}^{(2)}(\omega,k)=\frac{1}{c^2}\int d\omega^\prime \text{A}_\text{p}(\omega^\prime,k^\prime)\text{K}^{(2)}(\omega,\omega^\prime)\text{A}_\text{pr}(\omega-\omega^\prime,k-k^\prime),
\end{align}
where we have performed the analytic continuation $i\Omega_m\to\omega+i0^+$ for all Matsubara frequencies. The second-order interaction kernel reads explicitly
\begin{align}
    \text{K}^{(2)}(\omega,\omega^\prime)=i\omega\,\omega^\prime(\omega-\omega^\prime)Z^*R\,D(\omega^\prime).
\end{align}
The frequency prefactors $\omega$, $\omega^\prime$ and $(\omega-\omega^\prime)$ are required by gauge invariance, and they are ultimately absorbed by the gauge potentials when computing the response to electric fields $\text{E}(\omega)=i\omega\text{A}(\omega)/c$.
\subsection{Propagation effects}
The internal response computed through the effective-action approach, alone, is not sufficient for the ultimate goal of understanding the experimental results. Optical experiments use pulses traveling from outside to within the finite-sized sample and then outside again, and the propagation inside the material and at its interfaces affects the spectral content of the final response. We thus need a generalized Maxwell-Fresnel approach to treat propagation effects of the fields in presence of nonlinear currents.

We consider a material in the region of space $0<\text{x}<d$, with $d$ the sample thickness, characterized by a finite second-order nonlinear kernel $\text{K}^{(2)}(\omega_1,\omega_2)$. Let us consider a generic electromagnetic field $\text{A}(\omega,\text{x})$ traveling along $\text{x}$ and inducing a nonlinear current within the sample,
\begin{align}
    \text{J}^{(2)}(\omega,\text{x})=\int d\omega_1 d\omega_2\text{A}(\omega_1,\text{x})\text{K}^{(2)}(\omega_1,\omega_2)\text{A}(\omega_2,\text{x})\delta(\omega-\omega_1-\omega_2).
\end{align}
This current acts as a source in Maxwell's equations for the nonlinear signal,
\begin{align}\label{nlsys}
    \begin{cases}
    \partial_\text{x}^2\text{A}(\omega,\text{x})+\frac{\omega^2}{c^2}\text{A}(\omega,\text{x})=0 &(\text{x}<0,\, \text{x}>d),\\
    \partial_\text{x}^2\text{A}(\omega,\text{x})+\frac{n^2(\omega)\omega^2}{c^2}\text{A}(\omega,\text{x})=-\frac{4\pi}{c}\text{J}^{(2)}(\omega,\text{x}) &(0<\text{x}<d),
    \end{cases}
\end{align}
where the refractive index $n(\omega)$ already includes the effects of linear currents on the nonlinear signal. In principle, the equation within the sample can only be solved self-consistently due to the dependence on the gauge fields of the nonlinear current. Nonetheless, since nonlinear effects are usually smaller with respect to linear ones, it is possible to solve the equation with a perturbative approach. We introduce a fictitious small parameter $\eta$ such that we can redefine the kernel as $\text{K}^{(2)}(\omega_1,\omega_2)=\eta\tilde{\text{K}}^{(2)}(\omega_1,\omega_2)$ and the electromagnetic potential as $\text{A}(\omega,\text{x})=\sum_n\eta^n\text{A}^{[n]}(\omega,\text{x})$. The nonlinear current can thus be rewritten as
\begin{align}
    \text{J}^{(2)}(\omega,\text{x})=\sum_{n,m}\eta^{n+m+1}\int d\omega_1 d\omega_2 \text{A}^{[n]}(\omega_1,\text{x})\tilde{\text{K}}^{(2)}(\omega_1,\omega_2)\text{A}^{[m]}(\omega_2,\text{x})\delta(\omega-\omega_1-\omega_2).
\end{align}
We can then solve Eq.\ \eqref{nlsys} for every order of $\eta$. In the following we will consider the response up to first order, that describes the relevant process for a three-wave-mixing pump--probe experiment. 

At zero-th order the system reads
\begin{align}\label{nlsys0}
    \begin{cases}
    \partial_\text{x}^2\text{A}^{[0]}(\omega,\text{x})+\frac{\omega^2}{c^2}\text{A}^{[0]}(\omega,\text{x})=0 &(\text{x}<0,\, \text{x}>d),\\
    \partial_\text{x}^2\text{A}^{[0]}(\omega,\text{x})+\frac{n^2(\omega)\omega^2}{c^2}\text{A}^{[0]}(\omega,\text{x})=0 &(0<\text{x}<d),
    \end{cases}
\end{align}
which describes the propagation of the field inside the material in linear-response approximation. The solution in the region $0<\text{x}<d$ is
\begin{align}\label{sol0}
    \text{A}^{[0]}(\omega,\text{x})=\text{A}^t(\omega)e^{in(\omega)\omega\text{x}/c}+\text{A}^r(\omega)e^{-in(\omega)\omega\text{x}/c},
\end{align}
where $\text{A}^t(\omega)=\text{A}_\text{ext}(\omega)t(\omega)f(\omega)$ and $\text{A}^r(\omega)=\text{A}_\text{ext}(\omega)t(\omega)r(\omega)f(\omega)e^{2in(\omega)\omega d/c}$. Here, $\text{A}_\text{ext}(\omega)$ represents the spectrum of the external incident field, while
\begin{align}
    t(\omega)=\frac{2}{1+n(\omega)}\qquad r(\omega)=\frac{n(\omega)-1}{n(\omega)+1}\qquad f(\omega)=\frac{1}{1-r^2(\omega)e^{2in(\omega)\omega d/c}}
\end{align}
are, respectively, transmission, reflection, and Fabry-Pérot coefficients.

At first order in $\eta$ the system reads
\begin{align}\label{nlsys1}
    \begin{cases}
    \partial_\text{x}^2\text{A}^{[1]}(\omega,\text{x})+\frac{\omega^2}{c^2}\text{A}^{[1]}(\omega,\text{x})=0 &(\text{x}<0,\, \text{x}>d),\\
    \partial_\text{x}^2\text{A}^{[1]}(\omega,\text{x})+\frac{n^2(\omega)\omega^2}{c^2}\text{A}^{[1]}(\omega,\text{x})=-\frac{4\pi}{c}\text{J}_{[1]}^{(2)}(\omega,\text{x}) &(0<\text{x}<d),
    \end{cases}
\end{align}
where
\begin{align}\label{nlcurrz}
    \text{J}_{[1]}^{(2)}(\omega,\text{x})=\int d\omega_1 d\omega_2 \text{A}^{[0]}(\omega_1,\text{x})\text{K}^{(2)}(\omega_1,\omega_2)\text{A}^{[0]}(\omega_2,\text{x})\delta(\omega-\omega_1-\omega_2).
\end{align}
The solution of Eq.\ \eqref{nlsys1} in $0<\text{x}<d$ can be written as
\begin{align}\label{nl_sig_int}
    \text{A}^{[1]}(\omega,\text{x})=\text{A}^{u}(\omega,\text{x})+\text{B}(\omega)e^{in(\omega)\omega\text{x}/c}+\text{C}(\omega)e^{-in(\omega)\omega\text{x}/c},
\end{align}
where $\text{B}(\omega)$ and $\text C(\omega)$ are the coefficients of the forward- and backward-propagating fields respectively, determined by boundary conditions, while $\text{A}^{u}(\omega,\text{x})$ is a unique solution to the differential equation. In particular,
\begin{align}\label{un_sol}
    \text{A}^{u}(\omega,\text{x})=\frac{4\pi}{c}\frac{1}{2\pi}\int dk \frac{e^{ik\text{x}}}{k^2-n^2(\omega)\omega^2/c^2}\text{J}^{(2)}_{[1]}(\omega,k),
\end{align}
with $\text{J}^{(2)}_{[1]}(\omega,k)$ obtained from Eq.\ \eqref{nlcurrz} by insertion of the solution Eq.\ \eqref{sol0} as
\begin{align}
    \text{J}^{(2)}_{[1]}(\omega,k)&=\int_0^dd\text{x}\,\text{J}^{(2)}_{[1]}(\omega,\text{x})e^{-ik\text{x}}\nonumber\\
    &=\int d\omega_1 d\omega_2 \delta(\omega-\omega_1-\omega_2)\text{K}^{(2)}(\omega_1,\omega_2)\sum_{\alpha_1}^{\pm1}\sum_{\alpha_2}^{\pm1}\bigg[\text{A}_{\alpha_1}(\omega_1)\text{A}_{\alpha_2}(\omega_2)\frac{1-e^{-i(k-\alpha_1k_1-\alpha_2k_2)d}}{i(k-\alpha_1k_1-\alpha_2k_2)}\bigg],
\end{align}
where, to keep a compact notation, we defined $k_i=n(\omega_i)\omega_i/c$, $\text{A}_{+1}(\omega)=\text{A}^t(\omega)$ and $\text{A}_{-1}(\omega)=\text{A}^r(\omega)$. The integral in Eq.\ \eqref{un_sol} can be solved with the residue theorem, and the unique solution reads explicitly
\begin{align}\label{un_sol2}
    \text{A}^{u}(\omega,\text{x})&=\frac{4\pi}{c}\int d\omega_1d\omega_2 \delta(\omega-\omega_1-\omega_2)\text{K}^{(2)}(\omega_1,\omega_2)\sum_{\alpha_1}^{\pm1}\sum_{\alpha_2}^{\pm1}\bigg[\text{A}_{\alpha_1}(\omega_1)\text{A}_{\alpha_2}(\omega_2)\nonumber\\
    &\times\bigg(\frac{e^{i(\alpha_1{k}_1+\alpha_2{k}_2)\text{x}}}{(\alpha_1{k}_1+\alpha_2{k}_2)^2-\frac{n^2(\omega)\omega}{c}}+\frac{e^{in(\omega)\omega\text{x}/c}}{\frac{2n(\omega)\omega}{c}(\frac{n(\omega)\omega}{c}-\alpha_1{k}_1-\alpha_2{k}_2)}+\frac{e^{i(\alpha_1{k}_1+\alpha_2{k}_2)d}\,e^{-in(\omega)\omega(\text{x}-d)/c}}{\frac{2n(\omega)\omega}{c}(\frac{n(\omega)\omega}{c}+\alpha_1{k}_1+\alpha_2{k}_2)}\bigg)\bigg].
\end{align}
We now have the solution for the nonlinear field propagating within the material, Eq.\ \eqref{nl_sig_int}. Outside the sample, the differential equation \eqref{nlsys1} has solutions $\text{A}^{[1]}(\omega,\text{x}>d)=\text{B}^\prime(\omega)e^{i\omega\text{x}/c}$ and $\text{A}^{[1]}(\omega,\text{x}<0)=\text{C}^\prime(\omega)e^{-i\omega\text{x}/c}$. Applying continuity conditions for the fields and their derivatives at the boundaries, one obtains the Fresnel system for the coefficients $\text{B}(\omega)$, $\text{B}^\prime(\omega)$, $\text{C}(\omega)$, and $C^\prime(\omega)$ of the fields, that reads
\begin{align}
    \begin{cases}
        \text{C}^\prime(\omega)=\text{A}^u(\omega,0)+\text{B}(\omega)+\text{C}(\omega)\\
        -\frac{i\omega}{c}\text{C}^\prime(\omega)=\text{A}^{u\prime}(\omega,0)+\frac{in(\omega)\omega}{c}(\text{B}(\omega)-\text{C}(\omega))\\
        \text{B}^\prime(\omega)e^{i\omega d/c}=\text{A}^u(\omega,d)+\text{B}(\omega)e^{in(\omega)\omega d/c}+\text{C}(\omega)e^{-in(\omega)\omega d/c}\\
        \frac{i\omega}{c}\text{B}^\prime(\omega)e^{i\omega d/c}=\text{A}^{u\prime}(\omega,d)+\frac{in(\omega)\omega}{c}(\text{B}(\omega)e^{in(\omega)\omega d/c}-\text{C}(\omega)e^{-in(\omega)\omega d/c}),
    \end{cases}
\end{align}
where $\text{A}^{u\prime}(\omega,\text{x})=\partial_\text{x}\text{A}^u(\omega,\text{x})$. In a transmission experiment, the relevant quantity that is being measured is the field collected by the detector after the sample, so we are interested in solving the boundary problem for the coefficient $\text{B}^\prime(\omega)$, obtaining
\begin{align}
    \text{B}^\prime(\omega)=-\frac{f(\omega)}{(n(\omega)+1)^2}&\bigg[2n(\omega)e^{in(\omega)\omega d/c}\big(\text{A}^u(\omega,0)-\frac{ic}{\omega}\text{A}^{u\prime}(\omega,0)\big)\nonumber\\
    -&\big(n(\omega)+1)\big)\big(n(\omega)\text{A}^u(\omega,d)-\frac{ic}{\omega}\text{A}^{u\prime}(\omega,d)\big)\nonumber\\
    +&\big(n(\omega)-1\big)e^{2in(\omega)\omega d/c}\big(n(\omega)\text{A}^{u}(\omega,d)+\frac{ic}{\omega}\text{A}^{u\prime}(\omega,d)\big)\bigg].
\end{align}
Finally, using the explicit expression for the unique solution Eq.\ \eqref{un_sol2}, we obtain
\begin{align}\label{Bprime}
    \text{B}^\prime(\omega)&=\frac{4\pi}{\omega}\frac{f(\omega)}{n(\omega)+1}e^{in(\omega)\omega d/c}\int d\omega_1 d\omega_2\delta(\omega-\omega_1-\omega_2)\text{K}^{(2)}(\omega_1,\omega_2)\nonumber\\
    &\times\sum_{\alpha_1}^{\pm1}\sum_{\alpha_2}^{\pm1}\bigg[\text{A}_{\alpha_1}(\omega_1)\text{A}_{\alpha_2}(\omega_2)\bigg(\frac{e^{i(\alpha_1{k}_1+\alpha_2{k}_2-n(\omega)\omega/c)d}-1}{\alpha_1{k}_1+\alpha_2{k}_2-n(\omega)\omega/c}+r(\omega)\frac{e^{i(\alpha_1{k}_1+\alpha_2{k}_2+n(\omega)\omega/c)d}-1}{\alpha_1{k}_1+\alpha_2{k}_2+n(\omega)\omega/c}\bigg)\bigg].
\end{align}
The generated nonlinear signal detected outside is then $\text{A}^{[1]}(\omega,d^+)=\text{B}^{\prime}(\omega)e^{i\omega d/c}$.
\subsection{Time-resolved pump--probe signal}
In ultrafast pump--probe techniques one measures the variations in the transmitted probe field as a function of the pump--probe time delay $t_{pp}$. In other words, the outgoing field becomes a function of the time delay, $\text{A}^{[1]}(\omega,d^+)\to\text{A}^{[1]}(\omega,t_{pp},d^+)$. In the formalism described above, this can be easily implemented in the external field, as
\begin{align}
    \text{A}_\text{ext}(\omega)\to\text{A}_\text{ext}(\omega,t_{pp})=\text{A}_\text{pr}(\omega)+\text{A}_\text{p}(\omega)e^{-i\omega t_{pp}}.
\end{align}
On the other hand, the observation time is fixed. This corresponds to an integration over $\omega$ in the frequency domain. The measured quantity in time domain $\text{A}_\text{NL}(t_{pp})$ will thus read
\begin{align}
    \text{A}_\text{NL}(t_{pp})=\int\text{A}^{[1]}(\omega^\prime,t_{pp},d^+)d\omega^\prime.
\end{align}
The spectral content of the pump--probe response is then found with a Fourier transform to the frequency associated with the pump--probe time delay $\Omega$:
\begin{align}
    \text{A}_\text{NL}(\Omega)=\int e^{i\Omega t_{pp}}\text{A}_\text{NL}(t_{pp})dt_{pp}.
\end{align}
Explicitly, using Eq.\ \eqref{Bprime}, we find
\begin{align}
    \text{A}_\text{NL}(\Omega)&=\int d\omega^\prime\frac{2\pi}{\omega^\prime}t(\omega^\prime)f(\omega^\prime)e^{i(n(\omega^\prime)+1)\omega^\prime d/c}\sum_{\alpha_1}^{\pm1}\sum_{\alpha_2}^{\pm1}\bigg[\text{A}_{\text{p},\alpha_1}(\Omega)\text{K}^{(2)}(\omega^\prime,\Omega)\text{A}_{\text{pr,}\alpha_2}(\omega^\prime-\Omega)\nonumber\\
    &\times\bigg(\frac{e^{i(\alpha_1k_1+\alpha_2k_2-n(\omega^\prime)\omega^\prime/c)d}-1}{\alpha_1k_1+\alpha_2k_2-n(\omega^\prime)\omega^\prime/c}+r(\omega^\prime)\frac{e^{i(\alpha_1k_1+\alpha_2k_2+n(\omega^\prime)\omega^\prime/c)d}-1}{\alpha_1k_1+\alpha_2k_2+n(\omega^\prime)\omega^\prime/c}\bigg)\bigg],
\end{align}
where now $k_1=n(\Omega)\Omega/c$ and $k_2=n(\omega^\prime-\Omega)(\omega^\prime-\Omega)/c$. Shifting to the electric field,
\begin{align}\label{Enl}
        \text{E}_\text{NL}(\Omega)&=\int d\omega^\prime\frac{2\pi i}{c}t(\omega^\prime)f(\omega^\prime)e^{i(n(\omega^\prime)+1)\omega^\prime d/c}\sum_{\alpha_1}^{\pm1}\sum_{\alpha_2}^{\pm1}\bigg[\text{A}_{\text{p},\alpha_1}(\Omega)\text{K}^{(2)}(\omega^\prime,\Omega)\text{A}_{\text{pr},\alpha_2}(\omega^\prime-\Omega)\nonumber\\
    &\times\bigg(\frac{e^{i(\alpha_1k_1+\alpha_2k_2-n(\omega^\prime)\omega^\prime/c)d}-1}{\alpha_1k_1+\alpha_2k_2-n(\omega^\prime)\omega^\prime/c}+r(\omega^\prime)\frac{e^{i(\alpha_1k_1+\alpha_2k_2+n(\omega^\prime)\omega^\prime/c)d}-1}{\alpha_1k_1+\alpha_2k_2+n(\omega^\prime)\omega^\prime/c}\bigg)\bigg].
\end{align}
The structure of this expression is composed of three main elements: the kernel of the interaction $\text{K}^{(2)}(\omega^\prime,\Omega)$, that encodes the internal response of the system as discussed above; the fields $\text{A}^t$ or $\text{A}^r$, that contain the external pulses and describe their propagation within the material; the terms in the round bracket, that encode the phase-matching factors. The latter are sinc-like functions peaked at the frequency in which the phase-matching condition is realized, i.e., $\Omega$ that makes the denominators vanish. For example, for the term scaling with $\text{A}_{+1}\text{A}_{+1}=\text{A}^t\text{A}^t$, we have
\begin{align}\label{sinc1}
    \frac{e^{i(k_1+k_2-n(\omega^\prime)\omega^\prime/c)d}-1}{k_1+k_2-n(\omega^\prime)\omega^\prime/c}=id\,e^{i(k_1+k_2-n(\omega^\prime)\omega^\prime/c)d/2}\bigg[\frac{\sin\big((k_1+k_2-n(\omega^\prime)\omega^\prime/c)\frac{d}{2}\big)}{(k_1+k_2-n(\omega^\prime)\omega^\prime/c)\frac{d}{2}}\bigg].
\end{align}
For this term, the phase-matching condition reads $\frac{n(\Omega)\Omega}{c}+\frac{n(\omega^\prime-\Omega)(\omega^\prime-\Omega)}{c}-\frac{n(\omega^\prime)\omega^\prime}{c}=0$. Considering that the refractive index is a complex quantity, so that we can define the wavelength $\lambda(\omega)=1/\text{Re}[n(\omega)\omega/c]$ and the penetration depth $\delta=1/\text{Im}[n(\omega)\omega/c]$ of the pulses, we could also rewrite Eq.\ \eqref{sinc1} as
\begin{align}
    \frac{e^{i(k_1+k_2-n(\omega^\prime)\omega^\prime/c)d}-1}{k_1+k_2-n(\omega^\prime)\omega^\prime/c}=d\frac{e^{i\frac{d}{\lambda}}e^{-\frac{d}{\delta}}-1}{\frac{d}{\lambda}+i\frac{d}{\delta}}=id\,e^{i\frac{d}{2\lambda}}e^{-\frac{d}{2\delta}}\bigg[\frac{\sin\big(\frac{d}{2\lambda}+i\frac{d}{2\delta}\big)}{\frac{d}{2\lambda}+i\frac{d}{2\delta}}\bigg],
\end{align}
with $\lambda^{-1}=\lambda_1^{-1}+\lambda_2^{-1}-\lambda^{-1}(\omega^\prime)$ and $\delta^{-1}=\delta_1^{-1}+\delta_2^{-1}-\delta^{-1}(\omega^\prime)$. One can then understand that the width of the phase-matching factors is controlled by the sample thickness and the penetration depth: for $d\gg\lambda$ and $\delta\gg\lambda$, the sinc functions are narrow and the phase-matched frequency dominates the response, allowing the use of TP-RP for the measurement of the dispersion. 

Eq.\ \eqref{Enl} contains all contributions to the nonlinear signal from forward- and backward-propagating pulses, as encoded in the sum over the indices $\alpha_{1,2}$. In practice, to describe the experimental Signals 1 and 2, we can neglect all terms scaling with backward-propagating NIR fields ($\alpha_2=-1$ and terms in $r(\omega^\prime)$). Moreover, since the sample is thick compared to the wavelength of the pulses, we simulate the experimental spectra with an approximated expression of Eq.\ \eqref{Enl}, in which we set the Fabry-Pérot factors to 1 and cut all the oscillating exponential terms. Under this assumption, we write the forward-going fields as $\text{A}^t(\omega)=\text{A}_\text{ext}(\omega)t(\omega)$, using the incident fields for $\text{A}_\text{ext}(\omega)$. The backward-going fields are instead written as $\text{A}^r(\omega)=\text{A}^{b}(\omega)t(\omega)r(\omega)$, where $\text{A}^b(\omega)=\text{A}_\text{ext}(\omega)e^{-d/\delta(\omega)}$ takes into account the field absorption through the medium. In practice, for the pump field we replace $\text{A}^b(\omega)$ with the measured transmitted THz field. We then separate the signal given by a forward-propagating pump term ($\alpha_1=+1$),
\begin{align}\label{E1}
    \text{E}^{(1)}_\text{NL}(\Omega)=-\frac{2\pi i}{c}\int d\omega^\prime t(\omega^\prime)\frac{\text{A}_\text{p}^t(\Omega)\text{K}^{(2)}(\omega^\prime,\Omega)\text{A}_\text{pr}^t(\omega^\prime-\Omega)}{\frac{n(\Omega)\Omega}{c}+\frac{n(\omega^\prime-\Omega)(\omega^\prime-\Omega)}{c}-\frac{n(\omega^\prime)\omega^\prime}{c}},
\end{align}
which describes Signal 1, and the signal given by a backward-propagating pump ($\alpha_1=-1$),
\begin{align}\label{E2}
    \text{E}^{(2)}_\text{NL}(\Omega)=-\frac{2\pi i}{c}\int d\omega^\prime t(\omega^\prime)\frac{\text{A}_\text{p}^r(\Omega)\text{K}^{(2)}(\omega^\prime,\Omega)\text{A}_\text{pr}^t(\omega^\prime-\Omega)}{-\frac{n(\Omega)\Omega}{c}+\frac{n(\omega^\prime-\Omega)(\omega^\prime-\Omega)}{c}-\frac{n(\omega^\prime)\omega^\prime}{c}},
\end{align}
which describes Signal 2. Finally, considering the polarizations of the pulses and the well-separated energies between pump and probe, we can substitute $n(\Omega)\to n_\text{THz}(\Omega)$, $n(\omega^\prime-\Omega)\to n_e$, and $n(\omega^\prime)\to n_o$. Analogously, $t(\Omega)\to t_\text{THz}(\Omega)$ and $r(\Omega)\to r_\text{THz}(\Omega)$, contained in $\text{A}^{t,r}_{\text{p}}(\Omega)$; $t(\omega^\prime-\Omega)\to t_e(\omega^\prime-\Omega)$, contained in $\text{A}^{t}_{\text{pr}}(\Omega)$; and $t(\omega^\prime)\to t_o(\omega^\prime)$, so that one retrieves Eqs.\ \eqref{for_prop} and \eqref{back_prop} of the main text.
\clearpage
\setcounter{equation}{0}
\setcounter{figure}{0}
\setcounter{table}{0}

\section{Comparison between experimental and theoretical spectra}\label{comparison}
\begin{figure}[h]
    \centering
    \includegraphics[width=0.8\textwidth]{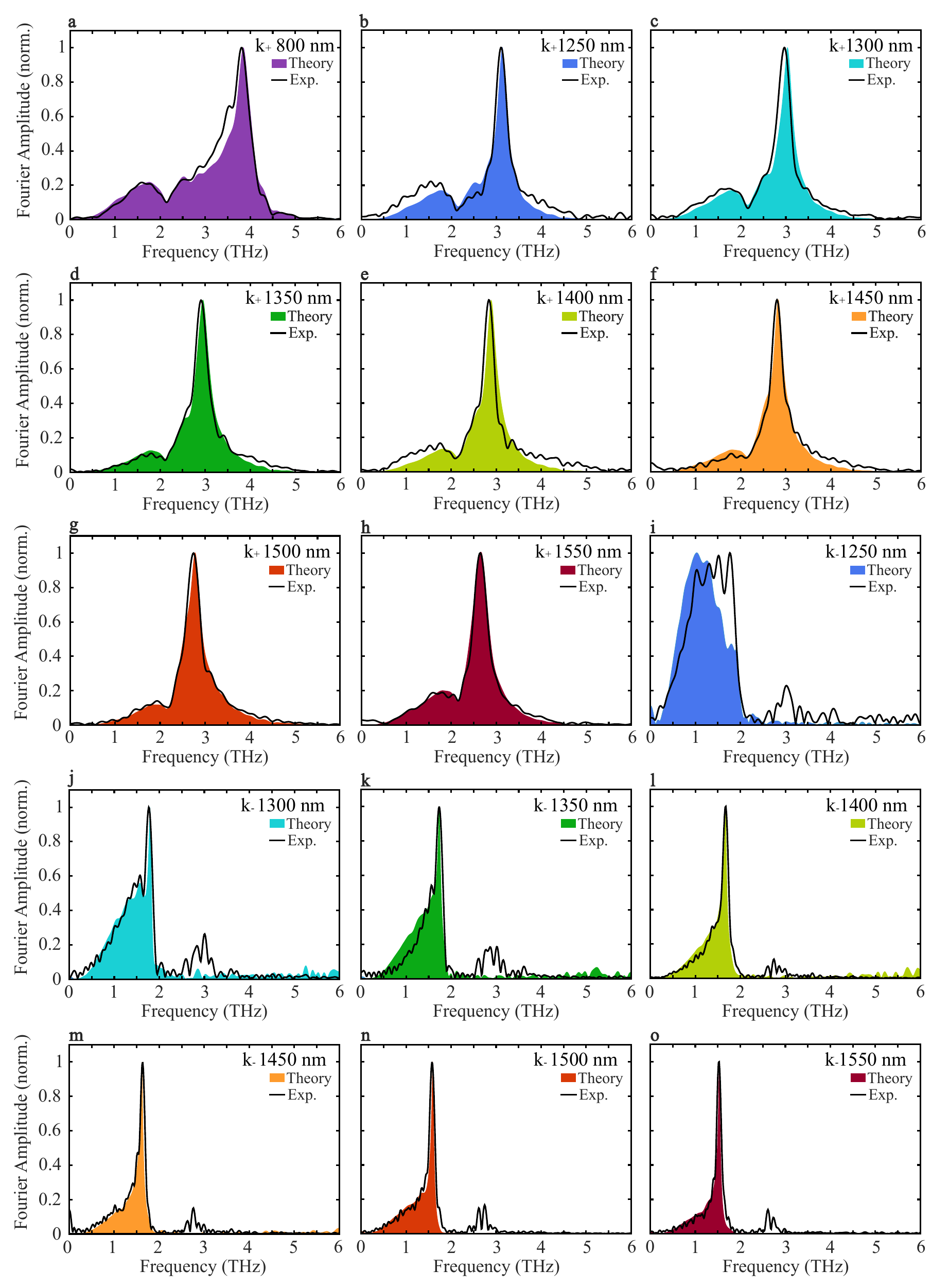}
    \caption{\textbf{Theoretical spectra for a complete dataset.} Calculated pump--probe response (colored areas) compared with the FFTs of the experimental signal (black curves) evaluated over the full time windows. Signal 1 is reproduced in panels a-h with probe wavelengths ranging from 800 nm to 1550 nm. Signal 2 is reproduced in panels i-o with probe wavelengths ranging from 1250 nm to 1550 nm. Values for $\varepsilon_\infty=22.47$, $\omega_\text{TO}/2\pi=4.44\text{ THz}$, and $\omega_\text{LO}/2\pi=5.94\text{ THz}$ are obtained from \cite{Knighton2018}. The phonon damping rate $\gamma$ is left as a free parameter in the calculations and changes between different spectra, as discussed in the main text.}
    \label{theorySim}
\end{figure}

In Fig.\ \ref{theorySim} we show the comparison between experimental spectra obtained with the FFT on the full time window (black solid lines), and calculated theoretical spectra obtained with Eqs.\ (\ref{for_prop}) and (\ref{back_prop}) of the main text, for one full dataset including all probe wavelengths. 

\end{appendices}
\end{document}